\documentclass[fleqn,usenatbib]{mnras}

\usepackage{newtxtext,newtxmath}

\usepackage[T1]{fontenc}

\DeclareRobustCommand{\VAN}[3]{#2}
\let\VANthebibliography\thebibliography
\def\thebibliography{\DeclareRobustCommand{\VAN}[3]{##3}\VANthebibliography}

\usepackage{graphicx}	
\graphicspath{ {./images/} }
\usepackage{amsmath}	
\usepackage{pifont}
\usepackage{mathtools}
\usepackage{placeins}
\usepackage{float}
\usepackage{bm}
    
\usepackage{nccmath}
\usepackage{color}
\usepackage{soul}
\usepackage{xcolor}

\usepackage{cleveref}
\crefname{figure}{Fig.}{Figs.}
\crefname{table}{Table}{Tables}

\definecolor{orange}{rgb}{1,0.5,0}
\definecolor{sred}{rgb}{.5,0,0}
\definecolor{sredsb}{rgb}{0.1,0,0}

\title[Radiation non-ideal MHD on a moving mesh]{Formation of protostars and the launching of stellar core outflows with moving-mesh radiation non-ideal magnetohydrodynamics}

\author[Mayer et al.]{%
Alexander C. Mayer$^{1,2}$\thanks{E-mail: amayer@mpa-garching.mpg.de}, Rüdiger Pakmor$^{1}$, Thorsten Naab$^{1}$, Oliver Zier$^{3}$, Alexei V. Ivlev$^{4,2}$, \newauthor Tommaso Grassi$^{4,2}$, Paola Caselli$^{4,2}$, Volker Springel$^{1}$
\vspace{0.1cm}\\%
$^{1}$Max-Planck-Institut für Astrophysik, Karl-Schwarzschild-Straße 1, 85741 Garching, Germany\\%
$^{2}$Excellence Cluster ORIGINS, Boltzmannstraße 2, 85748 Garching, Germany\\%
$^{3}$Center for Astrophysics | Harvard \& Smithsonian, 60 Garden St, Cambridge, MA 02138, USA\\
$^{4}$Max-Planck-Institut f\"ur Extraterrestrische Physik, Giessenbachstr. 1, 85748 Garching, Germany \\
}

\date{Accepted XXX. Received YYY; in original form ZZZ}

\pubyear{2025}

\begin{document}
 \label{firstpage}
\pagerange{\pageref{firstpage}--\pageref{lastpage}}
\maketitle

\begin{abstract}
We present an implementation of radiative transfer with flux-limited diffusion (FLD) for the moving-mesh code {\small AREPO} and use the method in a physical model for the formation of protostars with non-ideal radiation-magnetohydrodynamics (RMHD). We follow previous work in splitting the additional terms to the hydrodynamical equations arising from the inclusion of radiation into terms to be integrated explicitly and implicitly, as the diffusion and coupling terms would impose very restrictive timestep criteria. We validate the scheme with standard test problems for radiation diffusion, matter-gas coupling, and radiative shocks from the literature. Our implementation is compatible with local timestepping, which often presents problems for implicit schemes, and we found very good agreement with results obtained with global timesteps. We present an example application of the new implementation to the collapse of a $1\,{\rm M}_\odot$ molecular cloud core to a second Larson core modelled with radiation non-ideal magnetohydrodynamics. A high-velocity jet with v$_{\rm rad}> 10\, {\rm km\,s^{-1}}$ is self-consistently launched from the second core, nested within the first core, which produces a lower-velocity magnetorotational outflow. We observe magnetic field amplification up to more than $\vert \mathbf{B}\vert_{\rm max}>10^5$~G in the second core, which is surrounded by a small ($<0.5$~au) disk. This application demonstrates the robustness of our scheme in multi-scale and high-resolution simulations on arbitrary meshes and, as such, the model can be readily used for further simulations of protostar formation at high resolution.
\end{abstract}

\section{Introduction}
Radiation plays a central role in the formation of protostars \citep[][]{mckee_ostriker2007starformation,girichidis2020review,hennebelle_grudic2024imf,wuensch2024rt}. In the early phase of the collapse of a present-day molecular cloud core, the temperature is typically low ($\sim$~10~K) and the optical depth of the entire core is relatively small. As such, the core is approximately isothermal, despite the fact that the actual adiabatic index of the gas is much stiffer (from $\frac{5}{3}$ to $\frac{7}{5}$), but any radiation released by the collapse can escape easily. 

As the collapse proceeds, the optical depth in the centre rises along with the density, eventually leading to the formation of a \textit{first Larson core} (or hydrostatic core) as radiation from the collapse no longer escapes efficiently and the temperature rises \citep[cf. e.g. ][]{larson1969,masunaga_inutsuka2000secondcore,bate2010collapse,xu2021formationII}. Although the opacity drops when dust grains are evaporated\citep{semenov2003opacity}, the interior of the first hydrostatic core is essentially completely optically thick, dust and gas temperatures are tightly coupled, and the transport of radiation is diffusive. Further gravitational energy is released as the hydrostatic core accretes more mass and the radiation diffuses into the surroundings, such as the forming accretion disk. This escape of energy, along with the softening of the equation of state when H$_2$ starts to dissociate (at $\approx 2000$~K), eventually allows the collapse into a \textit{second Larson core} (or stellar core) \citep[cf. e.g. ][]{bate1998collapse,tomida2013rmhd,bate2014sphmhd,vaytet2018birth,ahmad2023protostar}. 

The temperature of protostellar disks is important in determining their susceptibility to fragmentation, as a hotter disk (or part of a disk) has more pressure support against gravitational fragmentation. This in turn affects whether a disk can split into a multiple star system, with implications for the initial mass function (IMF) \citep[cf. e.g.][]{bate2025imf}. Barotropic equations of state (where the pressure and thereby implicitly the temperature are set as function of density alone) are sometimes employed in simulations of protostar formation \citep[e.g. ][]{hennebelle2008magnetic,xu2021formationI} and often set to match the temperature evolution of the densest fluid element in the simulation and as such they do by construction underestimate the temperature in the surrounding region (as it is heated up precisely by the energy released from the collapse of the first material). Therefore, this simplification essentially always leads to colder disks than radiative calculations, which follow the temperature evolution self-consistently \citep{whitehouse2006rhd,bate2009imf,lebreuilly2021clumps}. 

The importance of magnetic fields in the star formation process has also been thoroughly established \citep{wurster_li2018review,pudritz_ray2019review,zhao2020review}, and hence many simulations of protostar and protostellar disk formation over the last two decades include ideal magnetohydrodynamics (MHD) \citep[e.g. ][]{seifried2013turbulence,bate2014sphmhd}. However, in turn, the need for non-ideal MHD effects to partially diffuse the magnetic field has been realized both as a way to prevent the `magnetic braking catastrophe' -- the non-formation of protostellar disks in ideal MHD due to efficient magnetic braking \citep{allen2003braking,galli2006braking} observed in some early MHD models -- and to solve the magnetic flux problem in the observed magnetic fields of young stellar objects. These fields are much smaller than inferred from the original molecular cloud, assuming  magnetic flux conservation \citep{zhao2020review}. 

The three main non-ideal MHD effects (Ohmic dissipation, ambipolar diffusion, and the Hall effect) emerge from the finite resistivity of the gas-dust mixture from which the star forms \citep{wardle1999conductivity}, which is increased strongly in molecular cloud cores because of their low ionization degree \citep{mckee_ostriker2007starformation,pineda2024ngc1333}. Therefore, a consistent chemical modelling is of importance as the local composition sets the strength of non-ideal MHD effects. The chemistry on the other hand needs accurate modelling of the local gas temperature, which requires radiative transfer in the partially or fully optically thick regimes of the protostellar disk and protostar itself. Hence, to improve the consistency of non-ideal MHD simulations of protostar formation, it is necessary to include a radiative transfer scheme. 

The first three-dimensional radiative simulations of protostar formation evolved beyond the formation of the stellar core were run by \cite{bate2010collapse}, where the launching of an outflow was observed without the inclusion of magnetic effects, purely through heating of the disk resulting from the energy released by the formation of the stellar core. \cite{tomida2013rmhd} performed the first non-ideal (including Ohmic resistivity) RMHD simulations and found outflow launching both from the first (in all magnetized models) and the second core (in their fast-rotating resistive model). Recently \cite{ahmad2023protostar,ahmad2025birth} have focused on following the early evolution of the protostar after the formation of the stellar core for a comparatively long time (more than half a year) with a resolution of $\sim 10^{-4}$~au.

Radiative transfer has been implemented in {\small AREPO} using the M1-closure \citep{kannan2019arepoRT, AREPO_RT_GPU}, a Monte-Carlo method \citep{smith2020mcrt}, and most recently a method using discrete ordinates \citep{ma2025idort}. There also exists a previous version of FLD \citep{bauer2015fld}, but only for post-processing on regular (adaptive mesh refinement) grids. Our aim is to provide a method that allows us to accurately describe optically thick regimes, with a relatively small computational cost, and that can be easily integrated with other modules of {\small AREPO}. In particular, we intend to use the radiation module to simulate the formation of protostars. We have, along with the FLD implementation, therefore built a model that includes an appropriate equation of state, opacities, and non-ideal MHD coefficients to simulate the entire process of cloud core collapse into a second Larson core. 

This work is structured as follows. In Section~\ref{sec:fld}, we first introduce the FLD approximation in terms of its governing equations and discuss its limitations. Next, the actual implementation is detailed in Section~\ref{sec:implementation}. We then perform various pre-existing tests from the literature to establish the validity of our scheme in Section~\ref{sec:tests}. Then the first application is presented in Section~\ref{sec:spherical_collapse} in the form of the collapse of a spherically symmetric molecular cloud core. The next Section~\ref{sec:nonideal_collapse} presents our full physical model for the formation of protostars with non-ideal RMHD and showcases results from the first simulation using the model. We conclude and give a small outlook for future work in Section~\ref{sec:conclusions}.

\section{The flux-limited diffusion approximation}
\label{sec:fld}

\subsection{Basic equations}
\label{subsec:basic_equation}

The starting point for gray (i.e., integrated over all frequencies) moment-based radiative transfer in the co-moving frame is described by the following equations for the evolution of the density $\rho$, the momentum density with the velocity $\bm{\mathrm{v}}$, the gas energy density $E = \rho u + \frac{1}{2}\rho \bm{\mathrm{v}}^2$, the radiation energy density $E_{\rm r}$, and the radiative flux $\bm{F_{\rm r}}$, respectively \citep{mihalas_mihalas1984book,turner_stone2001fld,whitehouse_bate2004sph_fld},
\begin{equation}
    \frac{\partial \rho}{\partial t} + \nabla \cdot (\rho \bm{\mathrm{v}}) = 0,
\end{equation}
\begin{equation}
    \frac{\partial (\rho \bm{\mathrm{v}})}{\partial t} + \nabla \cdot (\rho \bm{\mathrm{v}} \bm{\mathrm{v}}^{\rm T} + p\mathbb{I}) = \frac{\sigma_{\rm F}}{c} \bm{F}_{\rm r},
\end{equation}
\begin{equation}
    \frac{\partial E}{\partial t} + \nabla\cdot [\bm{\mathrm{v}} (E + p)] = - \sigma_{\rm P} (4 \pi B(T) - c E_{\rm r}),
\end{equation}
\begin{equation}
    \frac{\partial E_{\rm r}}{\partial t} + \nabla\cdot (\bm{\mathrm{v}} E_{\rm r}) = - \nabla\cdot\bm{F}_{\rm r}  - \mathbb{P}_{\rm r}:\nabla\bm{\mathrm{v}} + \sigma_{\rm P}(4\pi B(T) - c E_{\rm r}),
\end{equation}
\begin{equation}
    \frac{\partial \bm{F}_{\rm r}}{\partial t} + \nabla\cdot (\bm{\mathrm{v}} \bm{F}_{\rm r}^{\rm T}) = - c^2\nabla\cdot\mathbb{P}_{\rm r}  - (\bm{F}_{\rm r} \cdot \nabla)\bm{\mathrm{v}} - \sigma_{\rm F} c  \bm{F}_{\rm r},
\end{equation}
where $p$ is the thermal pressure, $\sigma_{\rm F}$ is the Flux opacity, $\sigma_{\rm P}$ is the Planck opacity (both with units of length$^{-1}$); $c$ is the speed of light, $B(T)$ is the Planck function and $\mathbb{P}_{\rm r}$ is the radiation pressure tensor.

Flux-limited diffusion (FLD) is the simplest method of moment-based radiative transfer, as only one moment of the equation of radiative transfer is considered. The radiative flux is assumed to always be in the direction opposite to the gradient of the radiation energy density,
\begin{equation}
    \bm{F}_{\rm r} = - \frac{c\lambda}{\sigma_{\rm R}}\nabla E_{\rm r}.
\end{equation}
$\sigma_{\rm F} = \sigma_{\rm R}$ is set in the FLD approximation, where $\sigma_{\rm R}$ is the Rosseland opacity\footnote{Planck- and Rosseland opacity result from different methods of weighting the contribution of different frequencies when integrating over all of them to obtain the grey approximation: $\sigma_{\rm P}~=~\frac{\int_{0}^{\infty}\sigma(\nu)B(\nu,T)d\nu}{\int_{0}^{\infty}B(\nu,T)d\nu}$, while $\sigma_{\rm R}~=~\frac{\int_{0}^{\infty}\sigma(\nu)^{-1} \frac{\partial B(\nu,T)}{\partial T}d\nu}{\int_{0}^{\infty}  \frac{\partial B(\nu,T)}{\partial T} d\nu}$, where $\nu$ denotes the frequency.}
In the above equation, $\lambda\equiv\lambda(R)$ is a function of the dimensionless quantity $R \equiv \frac{\vert\nabla E_{\rm r} \vert}{\sigma_{\rm R}E_{\rm r}}$ and is the so-called \textit{flux-limiter}, giving the FLD method its name. It serves to limit the propagation speed of radiation to $c$ in optically thin regions (where the transport is free-streaming), which is achieved by requiring that $\lambda(R) \xrightarrow[]{R \to \infty} R^{-1}$.  In the optically thick limit, the transport is diffusive, i.e, $\lambda(R) \xrightarrow[]{R \to 0} \frac{1}{3}$. 

There are multiple specific forms that have been used for $\lambda (R)$ in the literature. We have chosen the Minerbo flux limiter here, following the {\small RAMSES} implementation \citep{commercon2011fld},
\begin{equation}
    \lambda (R) = \begin{cases} \frac{2}{3 + \sqrt{9 + 12R^2}} \, ; \, 0 \leq R \leq \frac{3}{2}, \\ 
    \frac{1}{1 + R + \sqrt{1 + 2R}} \, ; \, \frac{3}{2} < R \leq \infty \quad .
    \end{cases}
\end{equation}
With this closure relation for the moments, the Euler equations with radiation treated in the co-moving frame take the form (neglecting gravity and magnetic terms):
\begin{equation}
    \frac{\partial \rho}{\partial t} + \nabla \cdot (\rho \bm{\mathrm{v}}) = 0,
    \label{eq:continuity}
\end{equation}
\begin{equation}
    \frac{\partial (\rho \bm{\mathrm{v}})}{\partial t} + \nabla \cdot (\rho \bm{\mathrm{v}} \bm{\mathrm{v}}^{\rm T} + p\mathbb{I}) = - \lambda \nabla E_{\rm r},
\end{equation}
\begin{equation}
    \frac{\partial E_{\rm tot}}{\partial t} + \nabla\cdot [\mathrm{v}(E_{\rm tot} + p)] = - \mathbb{P}_{\rm r}:\nabla \bm{\mathrm{v}} - \lambda \bm{\mathrm{v}} \nabla  E_{\rm r} + \nabla \cdot \left(\frac{c \lambda}{\rho \kappa_{\rm R}} \nabla E_{\rm r}\right),
    \label{eq:etot}
\end{equation}
\begin{equation}
    \frac{\partial E_{\rm r}}{\partial t} + \nabla\cdot (\bm{\mathrm{v}} E_{\rm r}) = - \mathbb{P}_{\rm r} \nabla : \bm{\mathrm{v}} + \nabla \cdot \left(\frac{c \lambda}{\rho \kappa_{\rm R}} \nabla E_{\rm r}\right) + \kappa_{\rm P} \rho c (a_{r} T^4 - E_r ),
    \label{eq:erad}
\end{equation}
with the \textit{total} energy density $E_{\rm tot} = E + E_r$, the definition $\kappa_{\rm i}~\equiv~\frac{\sigma_{\rm i}}{\rho}$~($i\in$\{P,R\}) for the Planck and Rosseland opacities\footnote{Note that these are called `absorption coefficients' in other contexts.}, and finally the radiation constant $a_{\rm r}$.  The most challenging terms in the above equations are those involving the speed of light, as their associated timescales are much smaller than those of ordinary hydrodynamics. These are the radiation diffusion term appearing both in equations \eqref{eq:etot} and \eqref{eq:erad}, and the matter-radiation coupling term in equation \eqref{eq:erad}. We describe the numerical methods used to integrate these equations in Section~\ref{sec:implementation}.

\subsection{Regime of applicability and limitations}

As its name implies, the FLD approximation is mainly useful in regimes where the transport of radiation is diffusive because the regions of interest are highly optically thick. Even in optically thin regions, the transport of radiation is treated as diffusive in the FLD approach, meaning that it has no preferential direction except away from high radiation densities. This means that there is no way to accurately model shadowing, captured in the two-moment M1 closure method, as radiation will simply diffuse around the obstacle to the other side \citep{gonzalez2007heracles}. However, the M1 closure is dominated by numerical diffusion at high optical depth per cell \citep{kannan2019arepoRT} and as such it is not well-suited for modelling the interior of protostars. In the specific context of protostellar disk formation, the need to accurately model both optically thin and thick regimes simultaneously has been addressed by utilizing hybrid schemes \citep{mignon-risse2021hybrid}.
\section{Implementation}
\label{sec:implementation}

\subsection{The moving-mesh code {\small AREPO}}
{\small AREPO} \citep[][]{springel2010arepo,pakmor2016gradients,weinberger2020arepo} solves the hydrodynamical equations using a finite-volume method on an unstructured and moving grid. This grid is constructed as the Voronoi tessellation of a number of mesh-generating points, which approximately move with the local fluid velocity to reduce advection errors and provide automatically adaptive resolution. For the simulation including non-ideal MHD presented in Section~\ref{sec:nonideal_collapse}, we utilize the MHD implementation of {\small AREPO} presented in \cite{pakmor2011mhd} \citep[using the 8-wave Powell divergence control technique,][]{powell1999divergence,pakmor2012powell} and that of non-ideal MHD as described by \cite{zier2024diffusion}.

\subsection{Integration of equations}
\label{subsec:integration}

The general approach of splitting the equations of FLD into parts to be integrated explicitly and implicitly is seen across implementations in various codes \citep{turner_stone2001fld,whitehouse_bate2004sph_fld,zhang2011castro,commercon2011fld}. Our method is specifically inspired by the implementation of \citet{commercon2011fld} in {\small RAMSES}, which splits the terms in equations~\eqref{eq:continuity}-\eqref{eq:erad} into parts that modify the hydrodynamical Riemann solver and operator-split the source terms, both of which are integrated explicitly, and an implicit solve for the coupling and diffusion terms. 

We report the relevant equations here for completeness. The first step is to rewrite the terms including the flux limiter as $\lambda = \frac{1}{3} + (\lambda - \frac{1}{3})$, where the first term represents the behaviour in the optically thick regime, and the second term encodes deviations from this behaviour (note that always $\lambda \leq \frac{1}{3}$). The contributions from the first term, not including the flux limiter, are collected to form the hyperbolic system with radiation (now with the magnetic terms in Heavyside-Lorentz units):
\begin{equation}
    \frac{\partial \bm{U}}{\partial t} + \nabla \cdot \bm{F} = 0,
\end{equation}
where:
\begin{equation}
    \bm{U} = \begin{pmatrix} \rho \\ \rho \bm{\mathrm{v}} \\ E_{\rm tot} \\ E_{\rm r} \\ \bm{B} \end{pmatrix} \quad , \quad\mathbf{F} = \begin{pmatrix} \rho \bm{\mathrm{v}} \\ \rho \bm{\mathrm{v}} \bm{\mathrm{v}}^{\rm T} + (p + \frac{1}{3} E_{\rm r}) - \bm{B} \bm{B}^{\rm T} \\ E_{\rm tot} \bm{\mathrm{v}} + (p + \frac{1}{3} E_{\rm r}) \bm{\mathrm{v}} - \bm{B}(\bm{\mathrm{v}} \cdot \bm{B}) \\ E_{\rm r} \bm{\mathrm{v}} \\ \bm{B}\bm{\mathrm{v}}^{\rm T} - \bm{\mathrm{v}}\bm{B}^{\rm T} \end{pmatrix}.
\end{equation}
Note that $p$ and $E_{\rm tot}$ here both include there magnetic contributions. The characteristic wave-speeds of the system are modified by the inclusion of radiation, which takes the shape of an extra thermal contribution (omitting the modification due to magnetic fields for clarity):
\begin{equation}
    c_{\rm s} = \sqrt{\gamma \frac{p}{\rho}} \quad \rightarrow \quad c^{\rm r}_{\rm s} = \sqrt{\gamma \frac{p}{\rho} + \frac{4}{9}\frac{E_{\rm r}}{\rho}},
\label{eq:wave-speeds}
\end{equation}
which in turn also increases the fast magnetosonic wave-speed as compared to the non-radiative system. 

\begin{equation}
S_{\lambda} =  \begin{pmatrix} 0 \\ -(\lambda - \frac{1}{3})\nabla E_{\rm r} \\ -(\lambda - \frac{1}{3})(\bm{\mathrm{v}}\nabla E_{\rm r} + \nabla : \bm{\mathrm{v}}) \\ -\lambda E_{\rm r} \nabla : \bm{\mathrm{v}} \end{pmatrix} \, ,
\end{equation}
where \citep[as ][]{commercon2011fld} we made the simplification that the radiative pressure tensor is isotropic with $\mathbb{P}_{\rm r} = \lambda E_{\rm r}$. We use the cell-centred gradients as calculated by {\small AREPO} for the standard hydrodynamic calculations (including limiters), and integrate the terms with second-order Strang-splitting. 

Note that these source terms are not conservative. An alternative method is to avoid splitting the flux-limiter and modify the wave-speeds in order to include the full dependence on $\lambda$ in the Riemann-solver, as in \cite{zhang2011castro}. However, as they point out, the dependence is in general not straightforward, and as such, the wave-speeds are only exact in certain regimes; but their method is by construction fully conservative. For the present application, the source terms are essentially always subdominant compared to the other radiation terms, and we therefore do not try to improve this method further here.

Finally, we come to the aforementioned challenging terms, i.e., including the speed of light. With the internal energy per unit mass $U$, the remaining parts to be integrated are obtained by subtracting \eqref{eq:erad} from \eqref{eq:etot}:
\begin{equation}
    \frac{\partial (\rho U)}{\partial t} = - \kappa_{\rm P}\rho c (a_{\rm r} T^4 - E_{\rm r}),
    \label{diff_coup_U}
\end{equation}
\begin{equation}
    \frac{\partial E_{\rm r}}{\partial t} - \nabla \cdot (\frac{c \lambda}{\kappa_{\rm R}\rho} \nabla E_{\rm r}) = + \kappa_{\rm P}\rho c (a_{\rm r} T^4 - E_{\rm r}).
    \label{diff_coup_Erad}
\end{equation}
The associated required timestep sizes to achieve stability in an explicit scheme are the matter-radiation coupling timescale
\begin{equation}
    \Delta t_{\rm coup} \sim  \frac{1}{\kappa_{\rm P}\rho c}\, {\rm min} \left( \frac{\rho U}{\vert a_{\rm r}T^4 - E_{\rm r}\vert} , \frac{E_{\rm r}}{\vert a_{\rm r}T^4 - E_{\rm r}\vert}\right),
    \label{eq:diffusion_timescale}
\end{equation}
and the radiative diffusion timescale
\begin{equation}
\Delta t_{\rm diff} \sim \frac{\kappa_{\rm R} \rho}{c \lambda} r^2 \, 
\label{eq:coupling_timescale}
\end{equation}
with the cell-size $r$. As mentioned above, the speed of light is much larger than any hydrodynamical speed in the system and its appearance as $\Delta t \propto c^{-1}$ in the explicit timestep criteria makes the use of an alternative to the explicit scheme a practical requirement\footnote{As a typical example in our application consider gas in a protostellar disk with density $\rho \sim 10^{-10}$~g~cm$^{-3}$, which has $\kappa_{\rm R} \sim 0.1$~cm$^2$~g$^{-1}$ and is resolved at $r < 0.1$~au; this leads to $\Delta t_{\rm diff} < 10^3$~s in contrast to the ordinary hydrodynamics timescale with $c_{\rm s} \sim 1$~km~s$^{-1}$, which results in $\Delta t_{\rm CFL} < 10^7$~s.}. 

However, using the (implicit) backward Euler integration of equations~\eqref{diff_coup_U} and \eqref{diff_coup_Erad} leads to an unconditionally stable scheme that is first-order convergent in time:
\begin{equation}
\label{implicit_internal}
    \frac{\rho_i U_i^{(n+1)} - \rho_i U_i^{(n)}}{\Delta t} = - \kappa^{(n+1)}_{{\rm P},i}\rho_i c \left[a_{\rm r} (T_i^{(n+1)})^4 - E^{(n+1)}_{\mathrm{r},i}\right],
\end{equation}
\begin{equation}
\label{implicit_radiation}
\begin{split}
    \frac{E^{(n+1)}_{\mathrm{r},i} - E^{(n)}_{\mathrm{r},i}}{\Delta t} - \nabla \cdot \left(\frac{c \lambda_i^{(n+1)}}{\kappa^{(n+1)}_{{\rm R},i}\rho_i} \nabla E^{(n+1)}_{\mathrm{r},i}\right) \\
    = \kappa^{(n+1)}_{{\rm P},i}\rho_i c \left[a_{\rm r} (T^{(n+1)})^4 - E^{(n+1)}_{\mathrm{r},i}\right].
\end{split}
\end{equation}
Here, superscripts $(n)$ and $(n+1)$ denote variables before and after this joint diffusion and coupling update, respectively. The spatial discretization of the diffusion term is described in Sect.~\ref{subsec:gradients}.

Equation \eqref{implicit_internal} contains the non-linear emission term $T^4$, which complicates solving the coupled system with an implicit scheme. \cite{commercon2011fld} use the linearization $(T^{(n+1)})^4\approx 4 (T^{(n)})^3 T^{(n+1)} - 3 (T^{(n)})^4$ on the right hand side of equations~\eqref{implicit_internal} and \eqref{implicit_radiation}, which is valid for small changes in temperature over the implicit update step. This approximation is equivalent to performing only a single step in the Newton iteration for the combined system of all $U_i$ and $E_{\mathrm{r},i}$ in the simulation. Whether this leads to an acceptable solution is ultimately dependent on the size of the timestep used, which here corresponds to the hydrodynamical timestep. In general, this timestep can be much larger than the cooling time of a cell. 

We follow the formulation of \cite{zhang2011castro} for the full iteration, which captures the behaviour of the non-linear emission term even for large timesteps. By taking the Schur complement, one can advance the system by a full step of the Newton iteration, first updating the radiation energies and then the internal energies. Dropping the cell-index $i$ for clarity, the update within a single iteration (here from iteration $k$ to $k+1$) is\footnote{Note that their definition of $\kappa_{\rm P/R}$ includes a factor of $\rho$ as compared to ours and that we multiplied the equation by the timestep $\Delta t$.}: 
\begin{equation}
\label{eq:calculation_radiation}
\begin{split}
[1 + \Delta t(1-\eta^{(k)})c\kappa_{\rm P}^{(k)}\rho]E^{(k+1)}_{\mathrm{r}} - \Delta t\nabla \cdot (\frac{c \lambda^{(0)}}{\kappa^{(k)}_{\rm R}\rho} \nabla E^{(k+1)}_{\rm r}) \\
= \Delta t(1-\eta^{(k)})c\kappa_{\rm P}^{(k)}\rho (T^{(k)})^4 + [E^{(0)}_{\mathrm{r}} - \eta^{(k)} \rho (U^{(k)} - U^{(0)})],
\end{split}
\end{equation}
with $\eta^{(k)} \equiv 1 - \frac{C_{\rm V}^{(k)}}{C_{\rm V}^{(k)} + 4 \Delta t c \kappa_{\rm P}^{(k)}a_{\rm r}(T^{(k)})^3}$. A superscript of $(0)$ denotes values before the iterations are started. 

Note in particular that we keep the slope limiter fixed to improve stability. $C_{\rm V} = \frac{{\rm d (\rho U)}}{{\rm d} T}\vert_{V = {\rm const.}}$ is the heat capacity at constant volume. Note that $C_{\rm V}$ \textit{can not simply be obtained by inverting the equation} $T = \frac{\rho U}{C_{\rm V}}$ (assuming here that one already has a way to obtain~$U$~from~$T$) and needs to be calculated independently, as pointed out by \cite{boley2007internalenergy}; rather, it should be seen as being defined such that the above relation between $U$ and $T$ holds \citep[cf. ][]{bate_keto2015ism}. The internal energy is correspondingly updated as:
\begin{equation}
U^{(k+1)} = \eta^{(k)}U^{(k)} + (1-\eta^{(k)})U^{(0)} - \Delta t (1 - \eta^{(k)}) c \kappa_{\rm P}^{(k)}[a_{\rm r}(T^{(k)})^4 - E^{(k+1)}_{\mathrm{r}}].
\end{equation}
As the diffusion term is a linear combination of $E^{(k+1)}_{\mathrm{r},j}$ terms (with non-zero contributions only from $j=i$ and neighbouring cells, see the next subsection), equation~\eqref{eq:calculation_radiation} for the update in a single iteration can be brought into the form of a matrix equation, which can then be solved via standard matrix methods (see the next subsection). 

Even for fixed heat capacity and opacity, the full equations are non-linear (and therefore cannot generally be treated in a single linear implicit solve) due to the flux-limiter and the quartic temperature dependence of the coupling term. While it is possible to neglect for our application the former (see the discussion in Sect.~\ref{subsec:test_streaming}), we found the latter to be relevant, since considering only the linearized coupling term can cause artificially high temperatures in optically thin shocked regions (e.g., at the edge of the pseudodisk in our application). 

As convergence in the temperature is our main concern for the Newton iterations, we choose to iterate until the condition 
\begin{equation}
\label{eq:stopping_criterion}
\vert[4 (T^{(k)})^3 T^{(k+1)} - 3 (T^{(k)})^4] - (T^{(k+1)})^4\vert  / (T^{(k+1)})^4 \leq 0.01
\end{equation}
is fulfilled for every active cell, as this generally implies that the non-linearity in the coupling term is sufficiently taken into account. We limit the number of Newton iterations to a fiducial value of 25, but this is, in practice, rarely necessary and in that case only due to a small number of cells failing to converge, without significantly affecting our results. Note that one can improve the accuracy further with a stricter stopping criterion (cf. Subsection~\ref{subsec:test_coupling}) and that our criterion only checks for a stall of the Newton iteration, rather than actual per-cell convergence to the solution. The current number of 25 iterations is likely higher than it needs to be; the number was determined as the number needed to achieve the accuracy in equation~\eqref{eq:stopping_criterion} in the aforementioned coupling test. In practical applications, the hydrodynamical timestep is much closer to the cooling time in the region of interest than in that particular test. If the iteration fails to converge after the first 5-10 iterations, this is in applications always due to individual cells in the background region (see Figure \ref{fig:scatterphases_nonideal} which demonstrates that there are no visible artifacts in any cells of interest in our application). For performance reasons, it is useful to identify if such a stall is happening and exit the iteration as the accuracy of the solution will not improve with further iterations. While all simulations presented in this work stop only after 25 iterations, we have found good results from also allowing an early exit from the Newton iterations after a minimum of 5 iterations have been completed and the maximum of the left-hand side of equation \eqref{eq:stopping_criterion} over all active cells does not decrease anymore as compared to the previous iteration. We will likely utilize this updated stopping criterion in future work.

\begin{figure}
    \centering
    \includegraphics[width=1.0\linewidth]{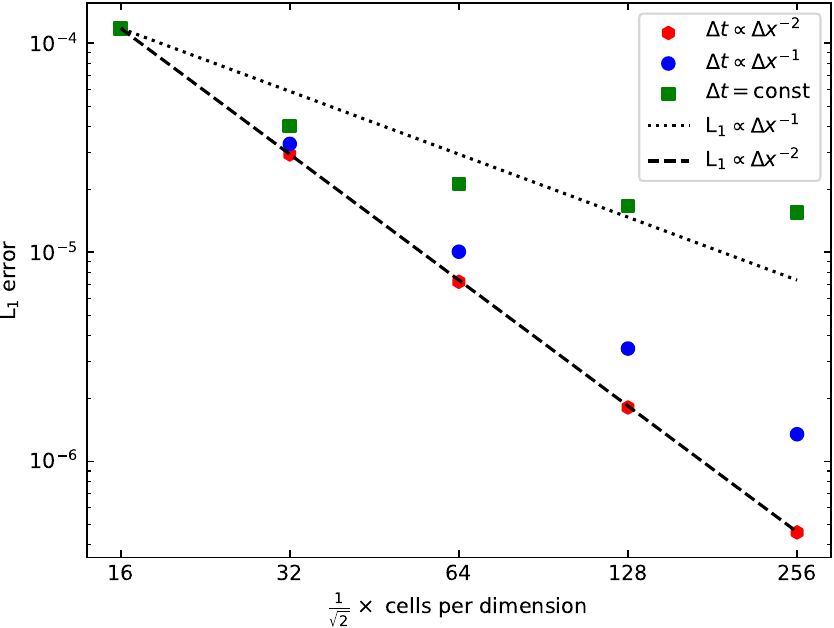}
    \caption{Test of convergence order for the diffusion of a Gaussian. We use a reference timestep of $\Delta t = 1.5625\times 10^{-14}$~s at the lowest resolution and test the effect of different scalings of the timestep with resolution: Quadratic decrease, linear decrease, and keeping the timestep the same. Second-order convergence is achieved for the quadratic decrease, while the time integration error begins to dominate at higher resolution in the other cases, degrading the convergence order. See Subsection~\ref{subsec:test_diffusion} for details.}
    \label{fig:test_diffusion}
\end{figure}

\subsection{Calculating gradients and solving the implicit equation}
\label{subsec:gradients}

We rely on the method of gradient estimation on interfaces presented in \cite{pakmor2016cr} for cosmic ray diffusion, which has been used in modified forms for subsequent implementations of other physics modules in {\small AREPO}, such as anisotropic thermal conduction \cite{talbot2024conduction} and non-ideal magnetohydrodynamics \citep{zier2024diffusion,zier2024hall}. The gradient of the energy density is estimated as a combination of the energy density of cells belonging to a given interface, by first finding an estimate at each corner. As the dependence on each of these cells is linear, it is suitable for an implicit solver. As discussed in \cite{pakmor2016cr}, a fallback is generally required for stability if the local geometry is such that the value of a corner needs to be extrapolated instead of interpolated from its surrounding cells. A simpler but less accurate method (which is used where a corner is sufficiently problematic) is to obtain a finite-difference estimate from the positions of the two mesh-generating points of the cells that form the interface. This always underestimates the actual normal gradient, but is second-order accurate if the mesh-generating points are in the positions of the centres of mass (which is generally approximately true as {\small AREPO} uses mesh-regularization motions to achieve this). For isotropic diffusion, we do not require the components parallel to the interface, and the normal component is generally accurately calculated even by the simple estimate. As shown in \cite{pakmor2016cr}, both methods give accurate results in test problems for diffusion. We use the same criteria for a `failed' corner (which, as such, requires the fallback) as in \cite{pakmor2016cr}. For the diffusion coefficient at an interface, we use the arithmetic mean of the two cells that constitute the interface. We calculate the flux-limiter on a per-cell level, and update it along with the calculation of gradients at the centres of mass of cells.

To treat the matrix system that needs to be solved at every Newton iteration, we make use of solvers from the {\small HYPRE} library \citep{falgout2002hypre}. Following \cite{pakmor2016cr}, we use the GMRES (Generalized Minimal Residual Method) solver, but here we always apply the algebraic multigrid preconditioner as the solver generally fails to converge without it; this could be related to the presence of source terms in our application, making the system stiffer than for cosmic ray diffusion. In this article, we use thresholds for convergence (measured by the relative $L_2$-norm of the residual) from $10^{-8}$ (tests) to $10^{-12}$ (applications). If no convergence is reached after 50 iterations with the {\small HYPRE} solver\footnote{Note the use of the term `iterations' both for those of the linear solver and the Newton iterations discussed above. The iterations of the linear solver are required at every Newton iteration.}, we still accept the result, which in every case we have encountered is at most a factor of $10^{2}$ above the desired convergence threshold and therefore still a good solution. The Newton iteration also still uses it as the input for the next step. 

Non-convergence appears to only happen when all particles are active, but is not overly rare where this is the case (even if global timesteps are used). It is not completely clear why this is the case, but the phenomenon of much worsened convergence on global synchronization points is also observed for the solvers presented in \citet{pakmor2016cr,talbot2024conduction}. This issue could be addressed in the future by using an adaptive threshold for the convergence of the linear solver, depending on, e.g., the fraction of active cells. There is a further subtlety in the measure of convergence because one can multiply individual rows of the matrix equation by arbitrary non-zero factors while not changing the solution, but only the weighting of these rows for the convergence. We have found that in the application to protostellar collapse, the weight contributed by the first core is eventually so much higher than everything else in the simulation (as the core has much higher radiation energy density) that the background can escape to unphysical values as its solution is no longer constrained by the solver. We therefore multiply every row by the volume of its corresponding cell, which appears to completely prevent this issue in the cases tested.

\subsection{Extension to local timesteps}
For higher computational efficiency, {\small AREPO} is typically run in a configuration that utilizes local timestepping, where cells are assigned individual timestep sizes (based on criteria such as the local CFL criterion) that are nested by factors of two. Cells are advanced using their individual timestep rather than the global minimum of all timesteps (the latter can however be done for some test problems), such that only a smaller number of cells is advanced for all but the largest (`global') steps. 

The use of local timestepping is particularly helpful to efficiently run simulations with a large hierarchy in timestep sizes, such as cosmological simulations of structure formation, or our application to protostellar core collapse. Thus, we would like to extend our scheme to make use of it as well. As noted in previous work \citep[][]{tomida2013rmhd}, the extension of implicit schemes that forego the explicit timestep criteria (equations \eqref{eq:coupling_timescale} and \eqref{eq:diffusion_timescale}) to local timesteps is less straightforward than for explicit schemes. This is ultimately because in the explicit hyperbolic system, the stability criterion given by the usual CFL criterion already ensures that information cannot travel farther than the boundary of the active region. But this is not the case in our implicit scheme, and radiation could, in principle, propagate through the entire simulation region in a single step. 

As a further technical complication, we do not have access to the full mesh on non-global synchronization points, and building it for every timestep would largely eliminate the advantage of local timestepping (as other computations, such as the flux-calculations, generally have a subdominant cost compared to the mesh construction). Our initial implementation chose the approach proposed in \cite{pakmor2016cr} \citep[and also adopted by ][]{talbot2024conduction}, where the boundary layer of active cells (which already has to be included in the mesh-construction for the explicit hydrodynamics) is included in the implicit solve. However, this turned out to be both inaccurate and unstable in test problems and applications in our case. This appears to arise from a combination of the presence of the coupling term and the generally strong degree of violation of the explicit diffusion criterion in the problem at hand. As radiation cannot reach cells outside of the active region plus boundary layer, it is trapped by this numerical boundary (generally radiation will mostly be emitted from hotter cells which are on smaller timesteps). Within the solver, this then couples back to the gas internal energy, typically resulting in hotter-than-physical cells in the active region. The resulting difference in thermal pressure from a cell that was included in the solve and one that was not can be so large that it triggers shocks at the boundary of timebins. This problem is also not solved on the global synchronization steps, as the gas-radiation coupling takes a finite amount of time, resulting in the timestep pattern being clearly present in the internal energies output from the simulation (despite the radiation energy appearing smooth, as expected for a diffusion problem). While this approach has the advantage of being fully conservative, it is ultimately clearly not appropriate for the present application\footnote{This issue is similar to that pointed out in section 3.3.5 of \cite{commercon2014adaptive} for the use of the conservative Neumann boundary condition between active and non-active cells: The flux determined at the interface between them is often not a meaningful estimate, as the radiation could have propagated much further out within the timestep.}.

Instead, we follow the implementations in \cite{whitehouse2005faster_fld} and \cite{commercon2014adaptive} and do not allow a change in the radiation energy of non-active cells in the implicit solver, and instead use them as fixed (Dirichlet) boundary conditions. This is done by including non-active cells on the right-hand side of the matrix equation as presented by \cite{commercon2014adaptive}; just with more complicated geometrical factors (which are, however, already calculated for the boundary cells, as for active cells). In the conservative approach used for the aforementioned other implicit diffusion solvers in {\small AREPO}, the matrix elements resulting from the diffusion term use the timestep of the face; that is, the lower timestep of the two cells that form the interface. In our case, this would `miss' part of the flux from the cell on a smaller timestep to that on the larger one, as the values in the latter were assumed fixed during the solve for the former. As such, both the coupling and diffusion terms in the row associated with a cell always need to use the timestep of the cell itself. 

This approach of fixed boundary conditions outside the active region is not energy-conserving, but it turns out to perform well in both test problems and our physical model of the collapse of a molecular cloud core (see below) when compared to global timesteps. As explained above, this is in stark contrast to the perhaps intuitively preferred conservative approach. As the timestep $\Delta t$ is reduced (keeping everything else constant), fluxes in a step are also reduced proportionally to $\Delta t$. As such, we can expect energy losses to converge to zero at first order in time, the same as the convergence order of the overall method. 

\begin{figure}
    \centering
    \includegraphics[width=1.0\linewidth]{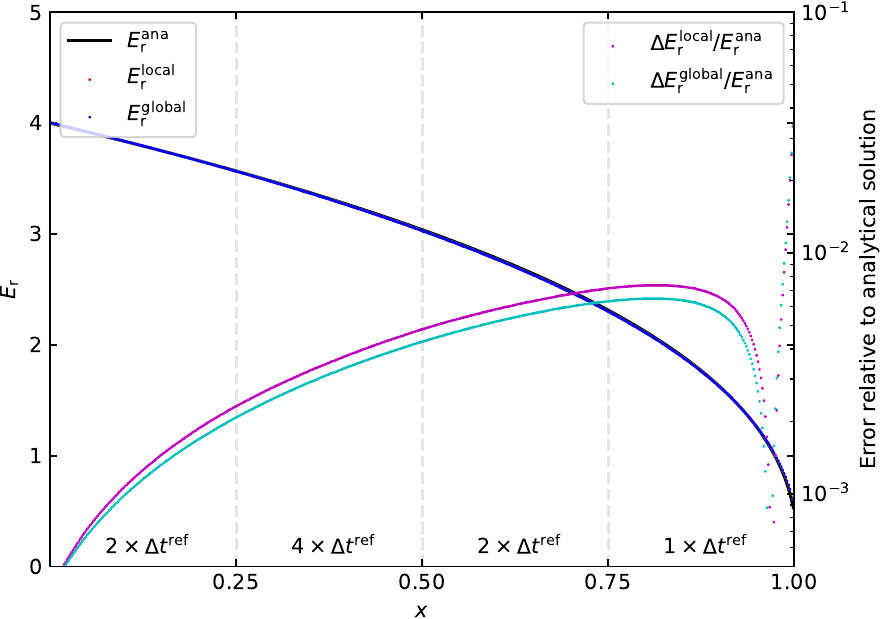}
    \caption{Radiation energy density as a function of position in the steady-state diffusion test, along with the analytical solution. We run both a simulation with a global timestep, where all cells are evolved with $\Delta t^{\rm ref} = 4.88281\times10^{-5}$, and one with local timesteps, where the timestep is set according to the values above the $x$-axis of the plot. Dashed grey lines denote boundaries between the timestepping regions. Both simulations show results very close to the analytical solution, and we do not observe any artifacts from local timesteps. See Subsection~\ref{subsec:test_diffusion} for details.}
    \label{fig:test_nonlinear_diffusion}
\end{figure}
\section{Test problems}
\label{sec:tests}

In this section, we present standard test problems as they have been shown previously in the literature. All tests are performed in two dimensions, as the determination of interface gradients works completely differently in one dimension and would, as such, not be representative of our actual application. On the other hand, the generalization of the scheme from two to three dimensions is straightforward, as described by \cite{pakmor2016cr}. In addition, {\small AREPO} does not have functionality for refinement in one dimension, and we also want to test the compatibility of our scheme with it.

\subsection{Radiation diffusion}
\label{subsec:test_diffusion}
We first only test pure radiation transport, with no coupling of the gas or hydrodynamics. We first examine the convergence of the scheme in a basic test problem, the diffusion of a Gaussian profile. We initialize the following profile on a periodic two-dimensional domain with side-length 2 at time $t = 5\times10^{-13} \, {\rm s}$:
\begin{equation}
    E_{\rm r}^{\rm ana} (r ,t) = 10^7 + \frac{10^5}{4\pi\chi t} \exp\left( -\frac{r^2}{4\chi t} \right) \, ,
\end{equation}
with $\chi = \frac{c}{3 \rho \kappa_{\rm R}}$, and in this test $\rho \kappa_{\rm R} = 1$. We enforce $\lambda = \frac{1}{3}$ for this calculation (the fully diffusive regime). The solution for any later time $t$ is also given by the above formula, and we calculate the L$_1$ norm of the error for different resolutions (all using a fixed mesh consisting of two Cartesian grids shifted relative to each other by one half of the grid-spacing) in Figure \ref{fig:test_diffusion}. 

Since our scheme is only first-order in time, one needs to reduce the timestep quadratically as a function of spatial resolution to achieve second-order convergence in both space and time. In line with earlier results which utilize the same scheme \citep{pakmor2016cr,talbot2024conduction}, the convergence is indeed second order in space if this is done. However, we also test the effect of only decreasing the time as $\Delta t \propto \Delta x$ and of keeping the timestep the same for all resolutions. The time integration error dominates over the spatial error at higher resolutions, making the convergence less than second order. In this problem, there is no explicit hydrodynamical CFL (only the product $\rho \kappa_{\rm R}$ is relevant in the problem, and as such the density can be rescaled at will), but in practice the timestep will be reduced linearly with resolution. In this specific test, this leads to convergence between first and second order.

Another test of pure radiation diffusion that has an analytical solution is presented by \cite{commercon2014adaptive}, specifically designed as a diffusion test with a steady-state solution that is independent of errors in the time integration. Here, the Rosseland opacity has a dependence on the radiation energy density via $\kappa_{\rm R} = 10^{10} E_{\rm r}^a$, with $a = {3}/{2}$. The box has a side length of 1, and the values of the radiation energy density are fixed at $E_{\rm r}^{\rm (0.0)} = 4.0$ and $E_{\rm r}^{\rm (1.0)} = 0.5$ at the left and right boundary, respectively. Initially, the radiation energy density values in the two halves of the box are set to the corresponding boundary values, but a steady state is quickly established, which has the analytical solution:
\begin{equation}
E_{\rm r}^{\rm ana} = \left[ \left( (E_{\rm r}^{\rm (1.0)})^{1 + a} - (E_{\rm r}^{\rm (0.0)})^{1+a}\right)x + \left( E_{\rm r}^{\rm (0.0)}\right)^{1+a}\right]^{\frac{1}{1+a}} \quad ,
\end{equation}
where $x$ is the coordinate within the box. The numerical results for both a simulation with local and with global timesteps, along with the analytical solution, are shown in Figure \ref{fig:test_nonlinear_diffusion} for a resolution of 512 cells in the $x$-direction. The simulation closely matches the analytical profile with an error profile similar to that found by \cite{commercon2014adaptive}, with the largest error appearing close to the right boundary, where the gradient in the radiation energy density is the largest.
\subsection{Limited propagation speed in optically thin medium}
\label{subsec:test_streaming}
\begin{figure}
    \centering
    \includegraphics[width=1.0\linewidth]{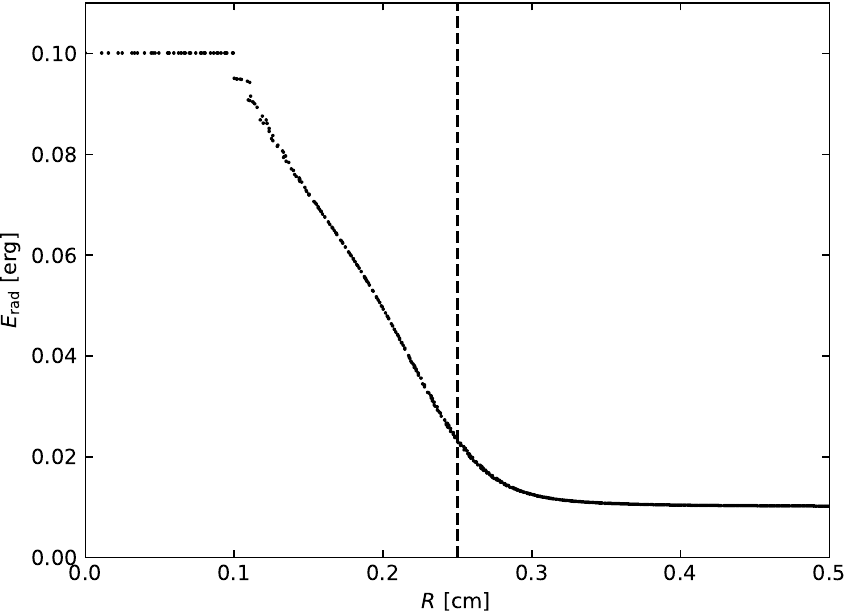}
    \caption{Propagation of a radiation pulse in an optically thin medium at time $t_{\rm end}=5\times10^{-12}$~s after the radiation is allowed to propagate out from the region with $R<1$~cm. Shown is the radiation energy density as a function of radius. The vertical line denotes the maximum propagation speed (as set by the speed of light). As required for physical consistency, the radiation front does not move away from the centre faster than $c$. See Subsection~\ref{subsec:test_streaming} for details.}
    \label{fig:test_limiter}
\end{figure}

As mentioned above, the flux-limiter $\lambda$ is introduced into the equations to prevent unphysical superluminal propagation of radiation. To verify that this limiting of the speed is actually enforced, we follow a set-up similar to \cite{whitehouse2005faster_fld}. 

To test the limiter, we set up a two-dimensional box of side length 1.0 filled with material that has $\rho\kappa_{\rm R} = 0.01$ cm$^{-1}$. Cells within a radius of 0.1 cm are set to an initial radiation energy density of $E_{\rm r}^{\rm inner} = 0.1$ erg cm$^{-3}$, while the cells outside are initialized to 0.2 erg cm$^{-3}$. The inner cells are fixed to $E_{\rm r}^{\rm inner}$ for the calculation. We run the simulation for a total time of $t_{\rm end}=5\times10^{-12}$~s, with a total of 64 time steps. Results for the radiation energy as a function of radius at $t_{\rm end}$ are displayed in Figure \ref{fig:test_limiter}. With $\chi = \frac{c}{3 \rho \kappa_{\rm R}}$, as defined before, and for $\lambda = \frac{1}{3}$ the characteristic scale the radiation would spread out over this time would be $\sqrt{\chi t_{\rm end}} \approx 0.7$ cm, i.e.~the radiation would have diffused over the entire range shown in the plot. The flux-limiter prevents this from happening by limiting the propagation speed to the speed of light. 

It has been consistently found that FLD shows strong diffusivity in this test \citep{whitehouse2005faster_fld,zhang2011castro}, as the physical interpretation suggests obtaining a sharp vertical line at the maximum propagation speed; however, the particular profile in our test is due to the two-dimensional setup. Since we keep the flux-limiter fixed during the Newton iterations, the propagation speed will not be limited correctly if only a single timestep is used, as found by \cite{whitehouse_bate2004sph_fld,whitehouse2005faster_fld}. However, since the region of interest in a simulation of protostar formation has $\lambda = 1/3$ everywhere and the accuracy of the FLD approximation is quite limited in optically thin regions in any case, this is not a problem for our application, and therefore we do not address this issue further.

\subsection{Matter-radiation coupling}
\label{subsec:test_coupling}

\begin{figure}
    \centering
    \includegraphics[width=1.0\linewidth]{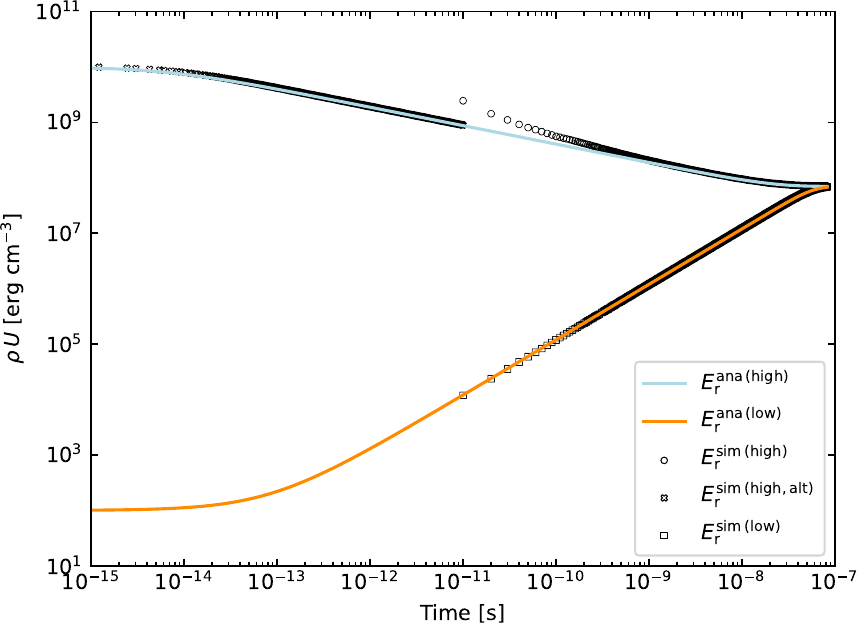}
    \caption{Test of the coupling terms between gas internal energy and radiation energy. Shown are results for the internal energy density for the two different initial conditions; initially above- (circles, crosses) and below-equilibrium (squares) temperature. The standard runs (circles, squares) use a timestep of $\Delta t =10^{-11}$~s. The simulation shown as crosses starts with the same initial conditions as the circles, but instead uses a smaller timestep ($\Delta t = 6.10352 \times10^{-16}$~s) and is only run to $t = 10^{-11}$~s. The analytical solutions for the case of high and low initial internal energy are shown as lines. The cooling is underestimated in the run depicted as circles due to the use of a large timestep size as compared to the cooling time, which can be addressed by taking smaller timesteps (crosses). See Subsection~\ref{subsec:test_coupling} for details.}
    \label{fig:test_coupling}
\end{figure}

\begin{figure*}
    \centering
    \includegraphics[width=0.45\linewidth]{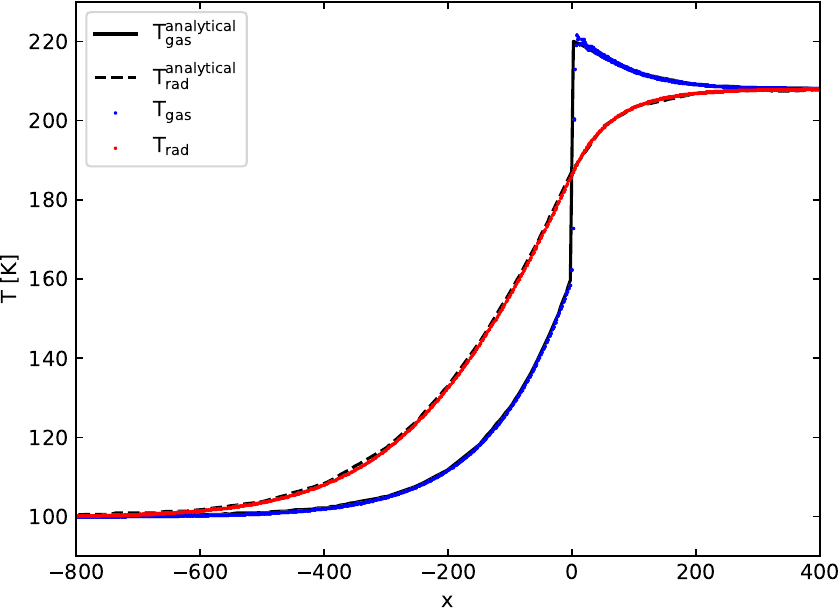}
    \includegraphics[width=0.45\linewidth]{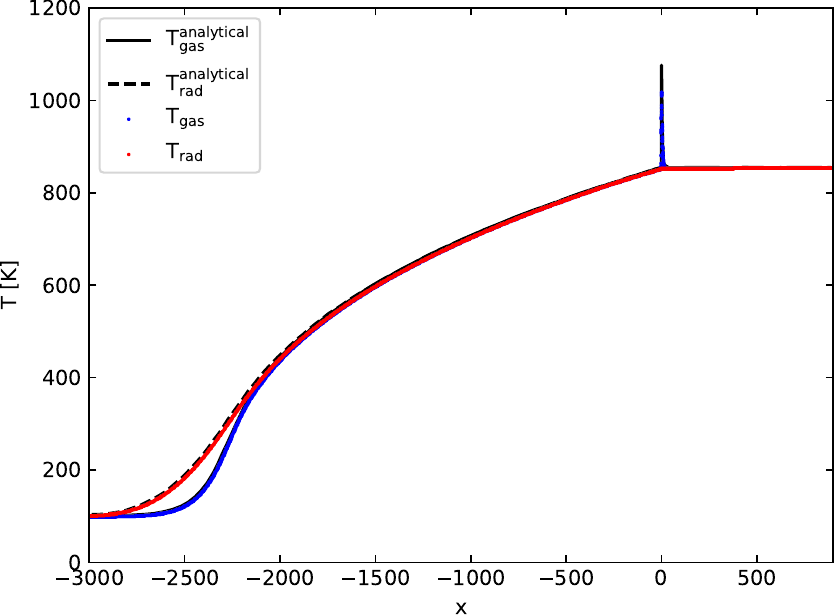}
    \caption{Gas and radiation temperature for the shocks with steady-state solution; subcritical case on the left, supercritical case on the right. We show the numerical results (points) along with the semi-analytical solutions (lines), the latter of which were extracted from Figure 7 of \protect\cite{commercon2014adaptive} and Figure 10 of \protect\cite{zhang2011castro}, respectively. The results are in good agreement with the expected solution. See Subsection~\ref{subsec:test_shocks} for details.}
    \label{fig:test_steady_shock}
\end{figure*}

\begin{figure*}
    \centering
    \includegraphics[width=0.45\linewidth]{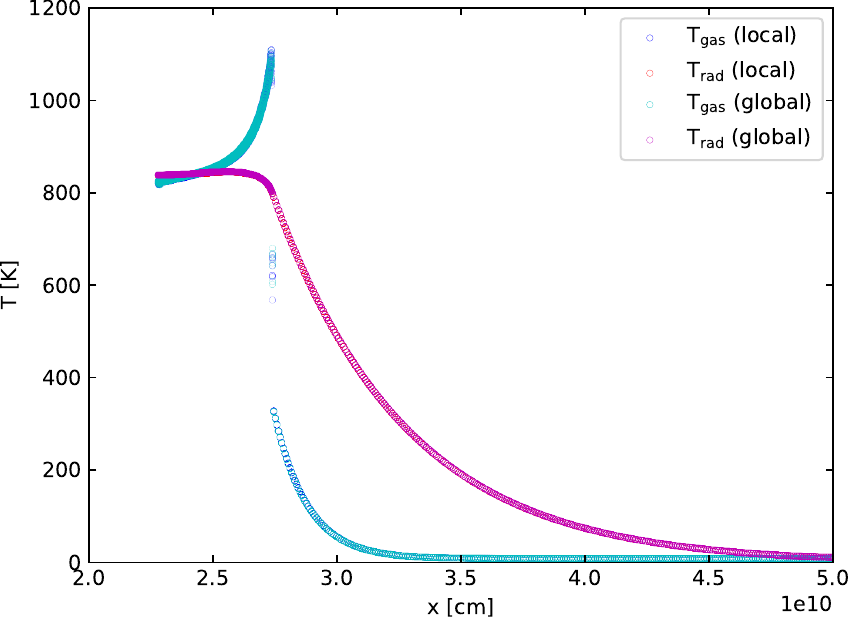}
    \includegraphics[width=0.45\linewidth]{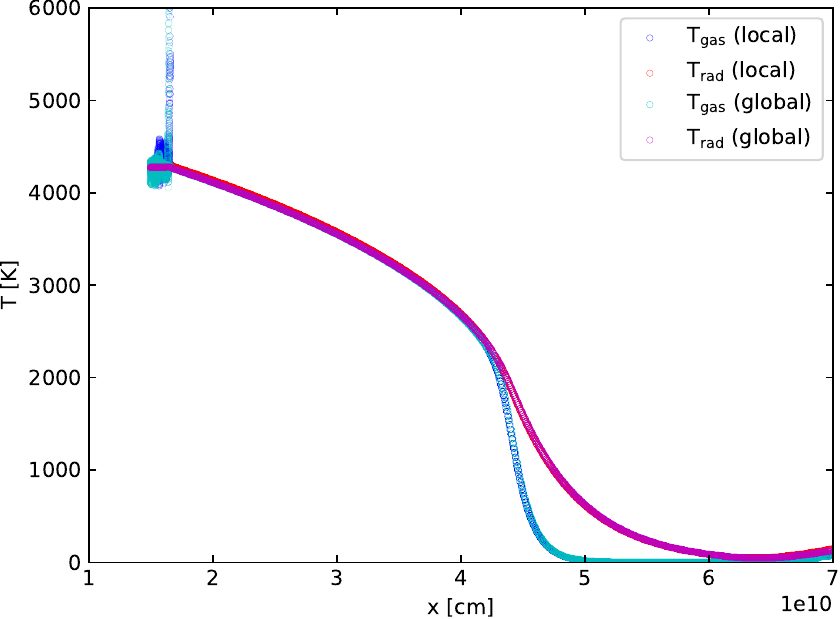}
    \caption{Gas and radiation temperature for the developed shocks with refinement; subcritical shock (left) and the supercritical shock (right). We show both results from simulations with local and global timesteps. The profiles are almost indistinguishable in the subcritical case and still very similar in the supercritical one. See Subsection~\ref{subsec:test_shocks} for details.}
    \label{fig:test_shocks}
\end{figure*}

To test the coupling terms between radiation and gas energy, we follow the setup of \citep{turner_stone2001fld,whitehouse_bate2004sph_fld,zhang2011castro,commercon2011fld}. The radiation energy density is set to $E_{\rm r}^{\rm (initial)} = 10^{12}$ erg cm$^{-3}$, and the gas temperature is initially out of equilibrium with the radiation. In one case, the gas internal energy density is set above the equilibrium temperature, to $\rho U_{\rm high} = 10^{10}$ erg cm$^{-3}$, and in the other case below, to $\rho U_{\rm low} = 10^{2}$ erg cm$^{-3}$. The density is set to $\rho = 10^{-7}$ g cm$^{-3}$, the Planck opacity is such that $\rho \kappa_{\rm P} = 4\times 10^{-8}$ cm$^{-3}$, and the mean molecular weight is $\mu = 0.6$ for $\gamma = \frac{5}{3}$. 

As the radiation energy is much larger than the gas energy in this test, it is almost constant during the calculation, and the analytical solution for the internal energy can be approximated by the solution to the ordinary differential equation
\begin{equation}
    \frac{{\rm d} (\rho U)}{{\rm d} t} = c \rho \kappa_{\rm P} \left[ E_{\rm r}^{\rm (initial)} - a_{\rm r} \left(\frac{\rho  U}{C_{\rm V}} \right)^4\right].
\end{equation}
The results of our code, run with a constant timestep of $10^{-11}$ s, along with the analytical solution for both initial conditions, are shown in Figure \ref{fig:test_coupling}. 

We observe that in both simulations the internal energy approaches the expected equilibrium, but at a reduced rate in the case of high initial gas energy. This matches previous results for this test, where the cooling rate is also underestimated in the setup with high initial temperature, even if an iteration is performed, in contrast to the simple linearization of the temperature term \citep{turner_stone2001fld,zhang2011castro}. Ultimately, the reason for the discrepancy from the analytical solution is that the timestep is much larger than the cooling time, and without iterating over the solution, the nonlinearity of the emission term is not fully taken into account. We have verified that convergence to the analytical solution with desired precision can be achieved even for such a large timestep by using a stricter convergence criterion, i.e., reducing the value in the right-hand side of \eqref{eq:stopping_criterion}. In our simulations of cloud-core collapse, we have found our convergence criterion to be adequate, considering the solid results of the shock tests discussed in the next subsection and the very similar results for local and global timesteps in the cloud-core collapse applications (cf. Section~\ref{sec:spherical_collapse}). 

\subsection{Radiative shocks}
\label{subsec:test_shocks}
\begin{figure*}
    \centering
    \includegraphics[width=1.0\linewidth]{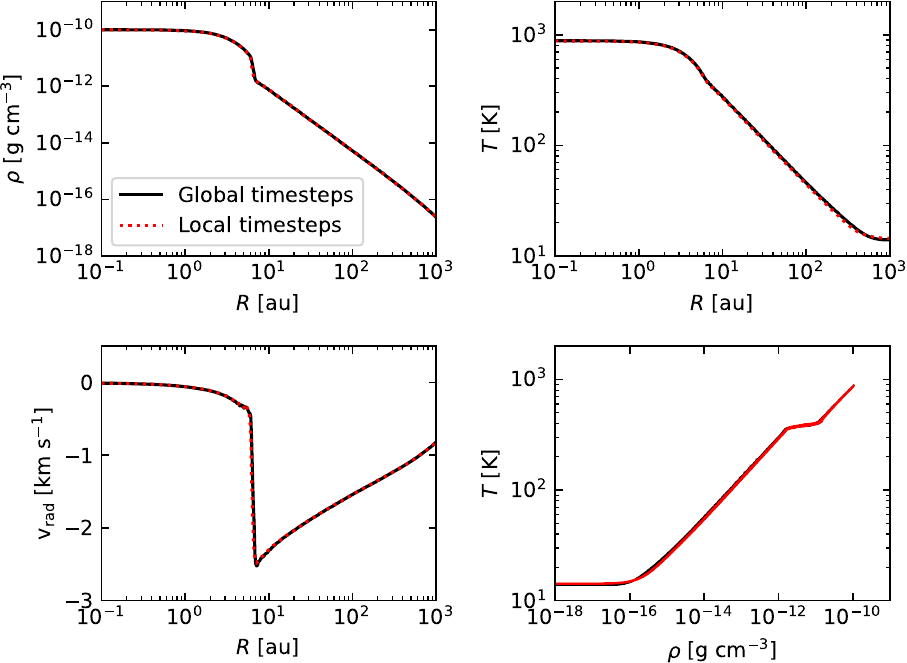}
    \caption{Radial profiles of density, temperature, and radial velocity profiles as well as the $\rho-T$ phase-space for the spherical setup with constant opacity and adiabatic index for both the simulations with global and with local timesteps. The differences are overall very small, with only slight deviations in the temperature profile in the outer regions (>~100~au). See Subsection~\ref{subsec:spherical_results} for details.}
    \label{fig:local_vs_global}
\end{figure*}
We now perform shock tests which probe all the components of our implementation simultaneously, as they involve the coupling of diffusion, gas-radiation coupling, and adiabatic compression/expansion. We run these tests using local timestepping, such that each cell evolves using its individual hydrodynamical timestep. Following previous work \citep[][]{turner_stone2001fld,whitehouse_bate2004sph_fld,commercon2011fld,zhang2011castro}, we simulate two different scenarios: Sub- and  supercritical shocks. In a supercritical shock, sufficient radiation is produced to heat upstream material (both in terms of gas and radiation temperature) nearly to the post-shock temperature, while this is not the case for the subcritical shock. 

First, we set up shocks with a steady-state solution taken from \cite{lorie_edwards2008shocks}, who also provided semi-analytical solutions. This setup was also studied by \cite{zhang2011castro} and \cite{commercon2014adaptive}. The subcritical shock is a Mach~2 shock on a domain ranging from -1000 cm to 500 cm, while the supercritical shock has a Mach number of 5 and the domain extends from - 4000 cm to 2000 cm. The initial left and right states are $\rho_{\rm L} = 5.45887\times 10^{-13}$~g~cm$^{-3}$, $T_{\rm L} =100$~K, v$_{\rm L}=2.35435\times 10^{5}$~cm~s$^{-1}$; $\rho_{\rm R} = 1.24794\times 10^{-12}$~g~cm$^{-3}$, $T_{\rm R} =207.757$~K, v$_{\rm R}=1.02987\times 10^{5}$~cm~s$^{-1}$ for the Mach~2 shock, and $\rho_{\rm L} = 5.45887\times 10^{-13}$~g~cm$^{-3}$, $T_{\rm L} =100$~K, v$_{\rm L}=5.88588\times 10^{5}$~cm~s$^{-1}$; $\rho_{\rm R} = 1.96405\times 10^{-12}$~g~cm$^{-3}$, $T_{\rm R} =855.720$~K, v$_{\rm R}=1.63592\times 10^{5}$~cm~s$^{-1}$ for the Mach~5 shock. In both cases the gas is characterized by constant $\gamma = \frac{5}{3}$, $\mu = 1.0$, $\rho \kappa_{\rm P} = 3.92664\times 10^{-5}$ cm$^{-1}$ and $\rho \kappa_{\rm R} = 0.848902$ cm$^{-1}$. 

In either case, we use periodic boundary conditions and extend the domain on both sides of this active region, and set all cells outside of the previously stated range to the initial values of either the left or right state, depending on their position. The full box is three times the size of the active region, extended by the entire length of the active region on either sides to prevent radiative diffusion from being effective over the periodic boundary. We set up the initial conditions of the shocks to approximately match the maximum effective resolution of \cite{zhang2011castro} (where AMR was used such that the resolution is lower in most of the domain), with 2048 cells in the $x$-direction for the subcritical shock ($\approx 683$ in the active region) and 16384 cells for the supercritical shock ($\approx 5461$ cells in the active region). For both shocks, we reduce the box in the $y$-direction (as the shock should be homogeneous in this direction) and use only 8 cells in this dimension. We specifically chose to run this setup at a high enough resolution not to use refinement, which causes some noise in the secondary dimension. We will test the effects of refinement in the other shock tests below. The results after 0.05 s for the subcritical shock and 0.04 s for the supercritical shock are displayed in Figure \ref{fig:test_steady_shock}, where we have shifted the numerical profile to the left by 20 cm in the subcritical and by 250 cm in the supercritical case. The profiles are in good agreement with the semi-analytical curves.

In a second pair of tests, we set up shocks that form from material flowing against a boundary from an initially homogeneous state, as originally described by \cite{ensman1994tests} but also tested by \cite{gonzalez2007heracles,commercon2011fld} and (in a slightly different setup) by \cite{whitehouse_bate2004sph_fld}. The size of the box is $7\times 10^{-10}$ cm, the initial density $\rho = 7.78 \times 10^{-10}$ g cm$^{-3}$ and the initial temperature $T=10$ K; the gas has $\gamma = \frac{5}{3}$, $\mu = 1.0$ and $\rho \kappa_{\rm P} = \rho \kappa_{\rm R} = 3.1\times 10^{-10}$ cm$^{-1}$. The gas is moving to the left with a velocity of v$_{\rm sub}= 6\times 10^5$ km s$^{-1}$ and v$_{\rm super}= 2\times 10^6$ km s$^{-1}$ in the sub- and supercritical case, respectively. To prevent numerical effects from a reflective boundary, we put the setup into the right half of our simulation domain (i.e., the active region) and add a mirrored setup in the left half (i.e.,~the full domain has size $1.4\times 10^{11}$ cm). There are 1024 cells along the $x$-direction in the active region and 8 in the $y$-direction, the box is shortened accordingly in the $y$-direction. 

Via refinement, cells are kept within a factor of 2 of the original mass (here $\approx 3.6\times10^6$~g; this is the standard refinement option in {\small AREPO}) and we additionally enforce a maximum area of cells of twice their initial size (this is used mostly to prevent `wall-heating' from becoming significant as all cells leave the region close to the boundary). With these two shocks we verify in a complete test problem the effect of refinement in our scheme, and to what extent the energy conservation is violated due to local timestepping. As there are no enforced boundary states (unlike in the steady-state shock), no energy should enter or leave the domain. 

The results for gas and radiation temperature are shown in Figure~\ref{fig:test_shocks}. When local timesteps are used, the total energy (radiation, internal, and kinetic) increases for both shock setups, where the relative increase is on the order of $10^{-3}$ for the supercritical shock, and $< 5 \times 10^{-4}$ for the subcritical shock. The profiles are only slightly changed by this energy non-conservation, and are in good agreement with the results with global timesteps \citep[as also found by ][]{commercon2011fld}. Deviations are larger in the supercritical case, which is a more difficult test in general and uses 5 timebins with local timestepping, as opposed to 4 in the subcritical case. There could be a connection between the noise in the internal energy to the left of the spike in the supercritical shock (but this is also present to a degree in the subcritical case). The profile is no longer uniform in the secondary $y$-direction, which is a result of small noise inherently produced by the refinement algorithm \citep[see also e.g.][]{zier2024diffusion}. As such, cells at a similar $x$-position can also be on different timesteps, enhancing deviations due to local timesteps. The main difference between our simulations and those of \cite{commercon2011fld} appears to be the height of the spike, but this is likely because the cell sizes in this region are smaller than that on their maximum AMR-level by a factor of a few, as we have checked that enforcing a similar minimum cell size causes the spike to lower to a value similar to theirs.
\section{Collapse of a spherically symmetric molecular cloud core}
\label{sec:spherical_collapse}

\subsection{Initial conditions and model}
\label{subsec:sperical_ics}
We first want to investigate potential differences between using local versus global timesteps in a toy setup. The initial conditions consist of a uniform non-rotating sphere of mass $M_{\rm core} = 1\, {\rm M}_\odot$ and radius $R_{\rm core} = 3000$~au, meant to serve as a simple model of a collapsing molecular cloud core \citep[specifically, this is a non-rotating and non-magnetized version of the initial conditions used in ][]{mayer2025isolated}. These parameters result in an initial density of $\approx 5.25\times10^{-18}$~g~cm$^{-3}$ and a free-fall time of $t_{\rm ff}\approx29$~Kyr. The initial temperature is set to $T_{\rm core} = 14 {\rm K}$, and we use a fixed molecular weight of $\mu = 2.381$ and an adiabatic index of $\gamma = \frac{5}{3}$, resulting in a ratio of thermal to gravitational energy of $\alpha_{\rm core}\approx 0.41$. 

The pressure in the background region (identified by a passive tracer in the initial conditions) is increased to prevent the boundary of the core from expanding while we enforce that these cells always keep the initial temperature to serve as a heat sink. To prevent boundary effects, we furthermore use periodic boundary conditions (with a domain of side-length 12000~au) for hydrodynamics, but the gravity is non-periodic. For simplicity, we assume a fixed opacity of $\kappa_{\rm P} = \kappa_{\rm R}=1$ g cm$^{-2}$ \citep[a lower limit to the typical opacity in the range of 100 to 1000~K, cf. e.g. ][]{semenov2003opacity}. We then run this setup both with global and local timesteps. For both simulations, we use the standard Lagrangian refinement strategy of {\small AREPO} and define a target mass of $3.33\times 10^{-7}$ M$_\odot$, where cells are split or merged if they deviate from this target by a factor of more than 2.

\subsection{Results}
\label{subsec:spherical_results}
After slightly more than a free-fall time ($t = 29.09$~Kyr), both the run with global and local timesteps have formed an optically thick and pressure-supported structure with increased density, a first (Larson) core, as seen in Figure \ref{fig:local_vs_global}. We selected the first snapshot of either simulation where the density has surpassed $10^{-10}$ g cm$^{-3}$. The output cadence of the run with local timesteps is smaller (as outputs are produced at global synchronization points only), but even so, the difference in time is only approximately a year. At this point, the simulation with local timesteps has developed a timestep hierarchy of 8 bins (resulting in a factor $2^7$ between the highest and lowest bin). This has already resulted in a speed-up of a factor of 4 as compared to the computational time required with global timesteps up to this point, but this advantage becomes even more severe  from this point on. 

We can see that the results are in very good agreement, with only small differences in the temperature of the outer regions. In both simulations, we can clearly identify the first core with a size of 6-7~au by a sharp drop in density and the fact that material just outside of it is moving inwards much more rapidly than just inside -- this marks the accretion shock onto the first core. The temperature changes more smoothly over the boundary of the hydrostatic core than the density,  as expected from the outward diffusion of radiation heating up the gas. This is shown by the $\rho$-$T$ phase-space as a flattening in the range of $10^{-12}-10^{-11}$~g~cm$^{-3}$ with temperature of 300-400~K.

\section{Collapse of a molecular cloud core with non-ideal RMHD}
\begin{figure}
    \centering
    \includegraphics[width=1.0\linewidth]{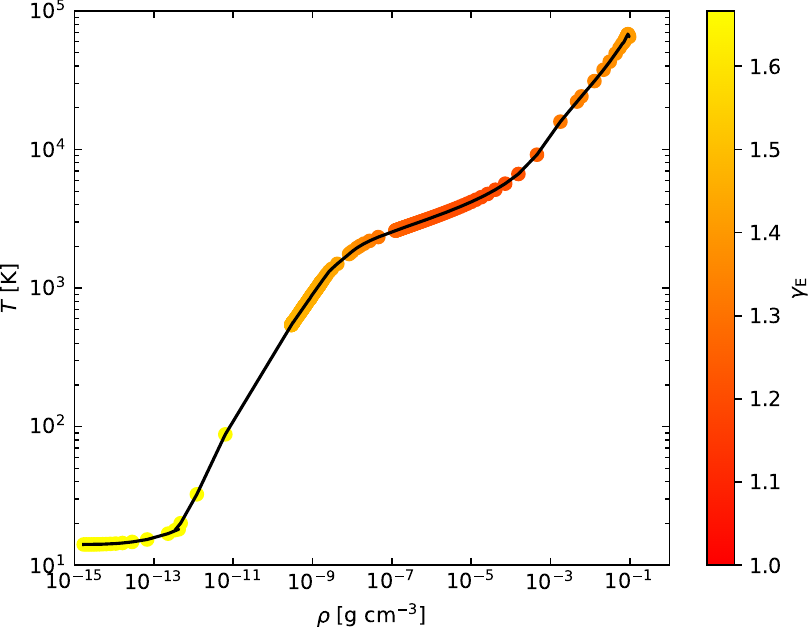}
    \caption{Temperature as a function of density for the gas element with maximum density at any given time along with the effective adiabatic index $\gamma_{\rm E} \equiv 1 + \frac{P}{\rho U}$ as encoded by the colour scale. Points show the underlying data points (all snapshots from $\rho_{\rm max}\approx10^{-15}$~g~cm$^{-3}$ on), connected by the black lines. Quickly after the cell rises above the background temperature it enters the optically thick regime where the evolution is essentially completely determined by the equation of state: The temperature rises more quickly as a function of density when $\gamma_{\rm E}$ is high and vice-versa. See Subsection~\ref{subsec:non_ideal_results} for details.}
    \label{fig:adiabatic_track}
\end{figure}

\begin{figure}
    \centering
    \includegraphics[width=1.0\linewidth]{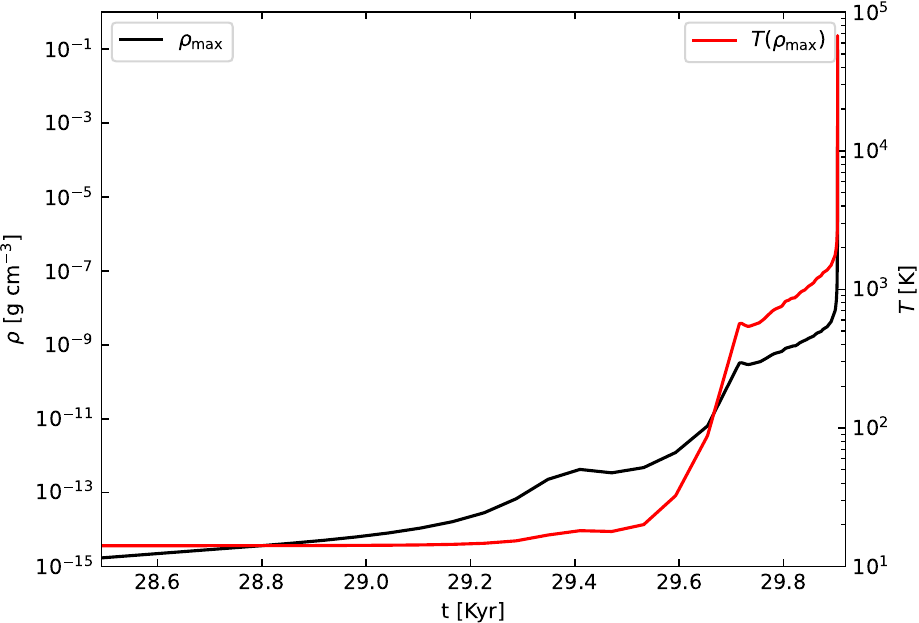}
    \caption{Evolution of the maximum density and temperature of the densest fluid element over time. The temperature starts rising as the central density becomes high enough to trap radiation, forming the first core. The second collapse $\rho > 10^{-9}$~g~cm$^{-3}$ proceeds very quickly as compared to the previous evolution of the first core. See Subsection~\ref{subsec:non_ideal_results} for details.}
    \label{fig:dens_over_time}
\end{figure}

\begin{figure*}
    \centering
    \includegraphics[width=0.45\linewidth]{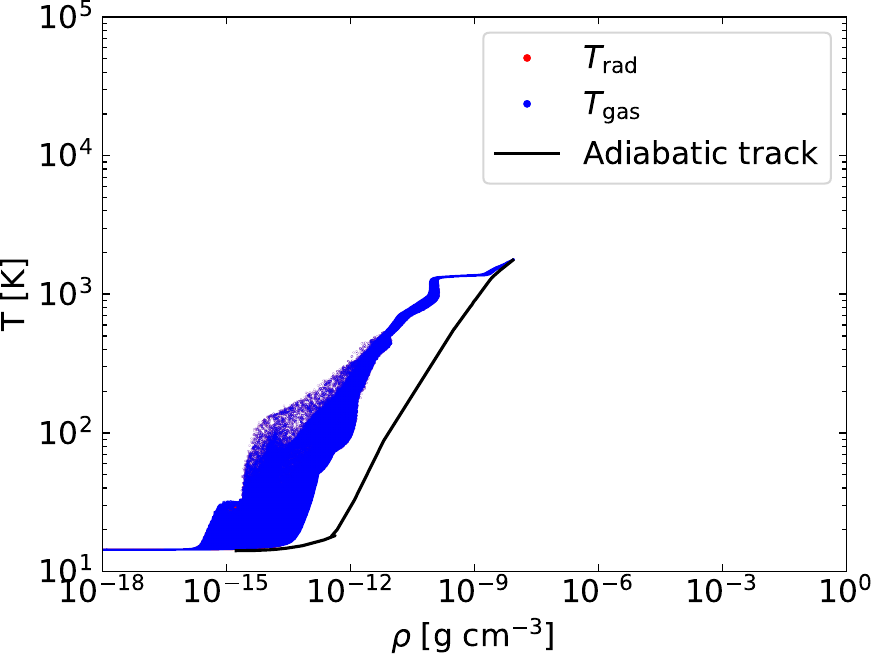}
    \includegraphics[width=0.45\linewidth]{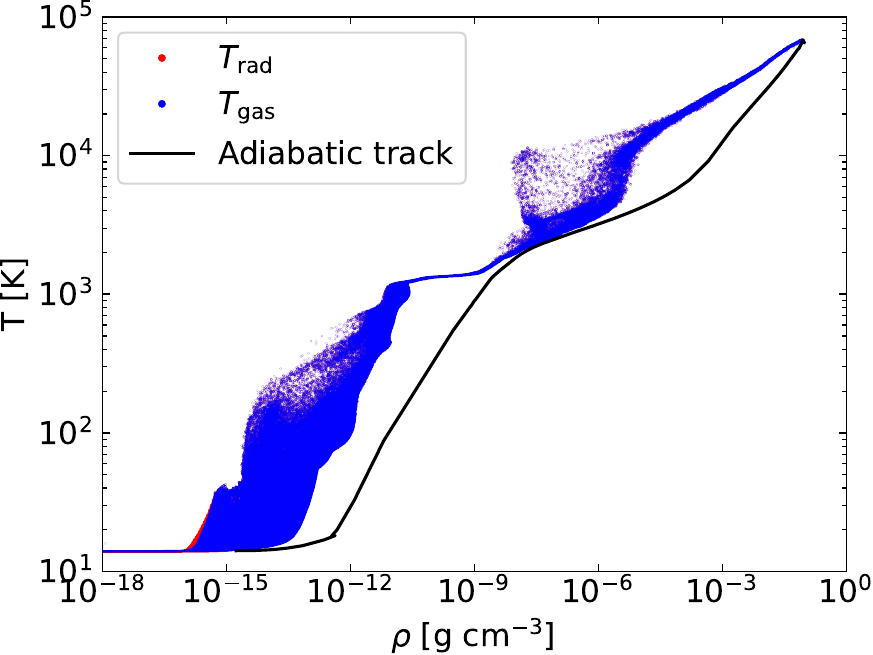}
    \caption{Full $\rho$-$T$ phase-space before the onset of the second collapse (left) and when the second core has formed and launched a fast outflow (right). The first and second core outflows can be seen as material that is much hotter than other gas at the same density. The adiabatic track from Figure \ref{fig:adiabatic_track} is added for comparison. Except for a regime of low opacity and $\gamma_{\rm E}$ at around $10^{-8}\,{\rm g\,cm^{-3}}$, the actual temperature is significantly above the adiabatic track. See Subsection~\ref{subsec:non_ideal_results} for details.}
    \label{fig:scatterphases_nonideal}
\end{figure*}
\label{sec:nonideal_collapse}

\subsection{Equation of state, opacity and non-ideal MHD model}

\begin{figure*}
    \centering
    \includegraphics[width=1.0\linewidth]{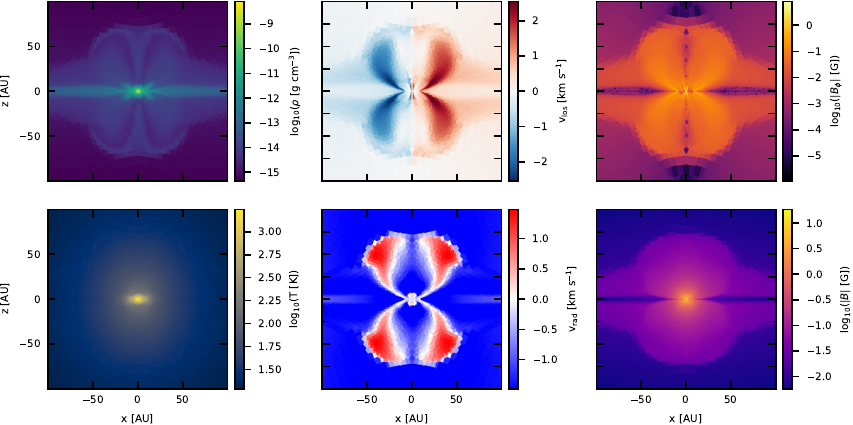}
    \includegraphics[width=1.0\linewidth]{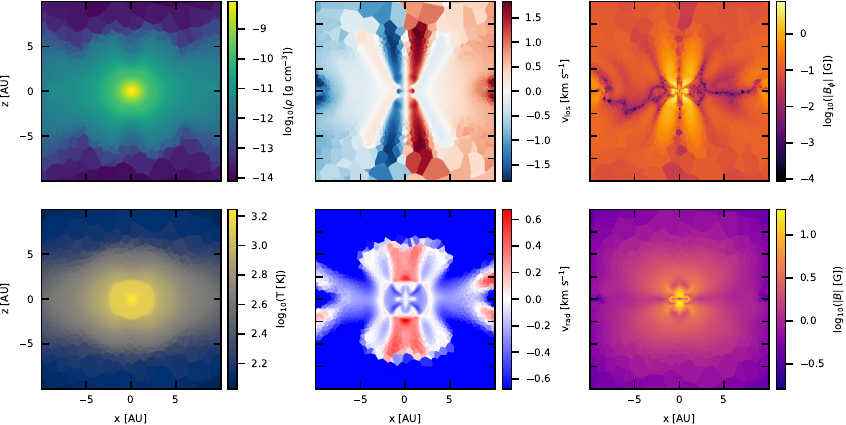}
    \caption{Several quantities on a slice in the $x$-$z$-plane at a time close to the onset of the second collapse, on the scale of the first-core outflow (6 top panels), and on the scale of the core itself (6 bottom panels). Top row, from left to right: Density, line-of-sight velocity, and magnitude of the azimuthal magnetic field. Bottom row, from left to right: Gas temperature, radial velocity, and magnetic field magnitude. The colour bar range corresponds to the full range present in each panel. The first core outflow is seen as material moving away from the centre in an `X' shape. In contrast to the second-core outflow (see next figure), it is not seen in the temperature, as it can cool efficiently. On the other hand, the magnetic field is significantly enhanced in the pseudo-disk and outflow cone. See Subsection~\ref{subsec:non_ideal_results} for details.}
    \label{fig:non_ideal_early}
\end{figure*}
\begin{figure*}
    \centering
    \includegraphics[width=1.0\linewidth]{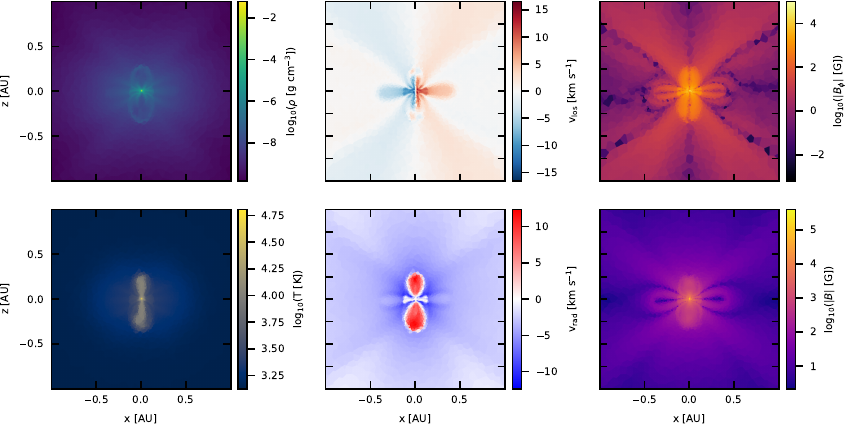}
    \includegraphics[width=1.0\linewidth]{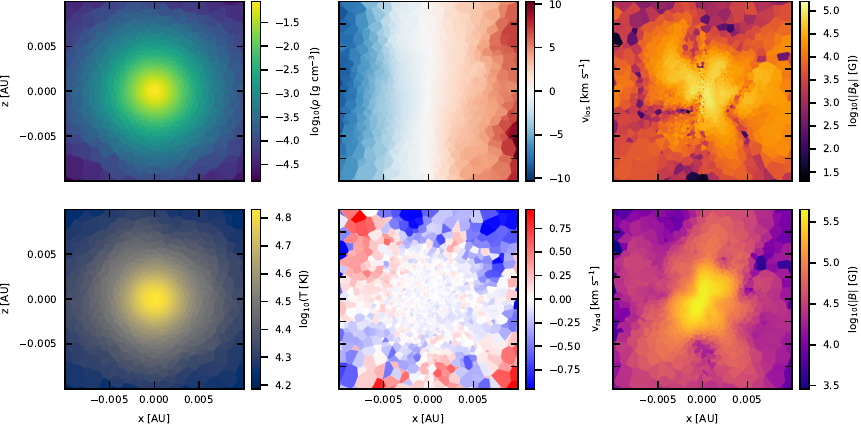}
    \caption{Same as Figure \ref{fig:non_ideal_early}, but at a later time, when the second core has been formed. The upper panels show the surroundings of the stellar core, and the lower panels the second core itself. The fast jet is seen most clearly in the radial velocity (in red), but can also clearly be identified in the temperature map in the upper panels. In the lower panels, we can see that the second core is very dense, \textit{not} rotating rapidly in contrast to its small surrounding disk, and strongly magnetized. See Subsection~\ref{subsec:non_ideal_results} for details.}
    \label{fig:non_ideal_late}
\end{figure*}
As mentioned in the introduction, the collapse of a molecular cloud core requires the modelling of a large range of physical regimes. One aspect is the thermodynamics, where we need to model regimes from a molecular cloud core composed of H$_2$ and neutral He to a gas mixture composed of fully ionized H and He with the corresponding electrons that ensure global charge neutrality, as well as the transition between these regimes over approximately 18 orders of magnitude in density ($\sim 10^{-19}-10^{-1}$ g cm$^{-3}$) and 4 orders of magnitude in temperature ($\sim 10-10^{5}$ K). To simulate protostar formation at low redshift, an equilibrium chemical model is generally utilized \citep{whitehouse2006rhd,tomida2013rmhd,vaytet2013multigroup}. While some of these studies use a tabulated equation of state, none of these tables are publicly available. As such, we chose to re-implement the model for chemistry and equation of state used by \cite{tomida2013rmhd} \citep[with the update from][]{tomida2015earlydisks} and described thoroughly therein. In brief, the model includes H$_2$, H, H$^+$, He, He$^+$, He$^{2+}$, and e$^-$ and assumes chemical equilibrium for everything but the ortho- to para-hydrogen ratio (which we set to a fixed value of 3:1, following their work). We assume hydrogen and helium nuclei abundances of respectively $X = 0.72$ and $Y=0.28$, i.e.,~we neglect the contribution of metals to the mean molecular weight, resulting in a slight decrease\footnote{The code used to generate the equation of state tables as well as the tables themselves can be found at \url{https://github.com/Seesternalge/protostellarEOS}.}. 

For our calculations, we tabulate the following four thermodynamic variables as a function of density and temperature: internal energy, pressure, heat capacity, and adiabatic index. The quantity 
\begin{equation}
    \gamma_{\rm E} \equiv 1 + \frac{p}{\rho U}
\end{equation}
is sometimes called the \textit{effective adiabatic index} and does not need to be tabulated independently (it is determined by $p$ and $U$). It is taken into account by the Riemann solver to set the correct relation between energy and pressure. $\gamma_{\rm E}$ is to be distinguished from the standard thermodynamical adiabatic index, which we \textit{do} provide in tabulated form:
\begin{equation}
\gamma_{\rm C} \equiv \left. \frac{\partial {\rm ln} P}{\partial {\rm ln} \rho} \right|_{S = {\rm const.}}
\end{equation}
and is also passed to the Riemann solver, but in this case to obtain the correct speeds for the wave-fan expansion as given by equation~\eqref{eq:wave-speeds}. As such, the timestep of the cell is also determined using $\gamma_{\rm C}$.
{\small AREPO} does not evolve the temperature itself, so the first step in determining the other quantities is to obtain the temperature from the internal energy density per unit mass. Since all other quantities are obtained from the table via bilinear interpolation, we determine the temperature via an inversion of this formula, as described in section 3.2.3 of \cite{vaidya2015eos}. 

The local opacity of the gas varies with density and temperature. For the dust-dominated opacities, we use the publicly available code that accompanies the study by \cite{semenov2003opacity}\footnote{\url{https://www2.mpia-hd.mpg.de/homes/semenov/Opacities/opacities.html}}. We choose the option of composite aggregate dust grains, as done by \cite{tomida2013rmhd}. At a temperature of 1500 K \cite[the same value used by][]{vaytet2013multigroup}, we switch to the gas opacities provided along with \cite{malygin2014opacity}\footnote{\url{https://vizier.cds.unistra.fr/viz-bin/VizieR?-source=J/A+A/568/A91}}. Note that, as in \cite{vaytet2013multigroup}, we are also including portions of the tables from \cite{semenov2003opacity} which represent gas opacities. The dust opacities are tabulated as functions of temperature and density, while the gas opacities use temperature and pressure, but this makes no real difference, as the pressure is fully determined by temperature and density via the equation of state. In a narrow region around the transition temperature, we blend the values from the two tables together to ensure the opacities are continuous functions of temperature. 

For the non-ideal MHD coefficients, we use {\small NICIL} \citep{wurster2016nicil,wurster2021nicil2.0} (version 2.1.1), as in our previous work \citep{mayer2025isolated,mayer2025zooms}. The tables are produced assuming a global cosmic-ray ionization rate of $\zeta = 10^{-16} {\rm s}^{-1}$, and a grain-size distribution with grains divided into 5 size bins distributed as a truncated MRN \citep{mathis1977mrn} profile from $a_{\rm min} = 0.1 \mu {\rm m}$ to $a_{\rm max} = 0.25 \mu {\rm m}$. We do not include the Hall effect, for simplicity, so that the chemical model simulated corresponds to the 16mnr00N run in \cite{mayer2025isolated}.

\subsection{Initial conditions and resolution}
The initial conditions for the initial core mass, radius, and temperature are identical to those stated in Subsection~\ref{subsec:sperical_ics}\footnote{This results in a very slight change in the ratio of thermal to gravitational energy as the initial mean molecular weight is a bit lower here, $\mu_{\rm init}\approx 2.324$.}, but here we add an initial magnetic field of $B_{\rm core} = 129$ $\mu$G (corresponding to a dimensionless mass-to-flux ratio of $\mu_{\rm core}=5$) and solid-body rotation with angular speed $\Omega_{\rm core} = 1.48\times 10^{-13}$ s$^{-1}$ (resulting in a ratio of rotational to gravitational energy of $\beta_{\rm core} = 0.005$). These are, except for the explicit setting of an initial temperature, identical to the initial conditions in \citep{mayer2025isolated} (this previous calculation used an effective equation of state). We again use a target mass of $3.33\times 10^{-7}$ M$_\odot$, but we impose a minimum volume of $4\times 10^{-12}$ au$^3$ (minimum cell-radius of $\approx 10^{-4}$ au) to speed up the calculation in the second core.

\subsection{Results}
\label{subsec:non_ideal_results}
As for the spherically symmetric collapse, we expect the formation of a first core, but as the collapse can then proceed further when molecular hydrogen is dissociated, finally, a second core should form. In agreement with previous studies \citep[e.g.][]{tomida2013rmhd,vaytet2018birth}, the temperature evolution of the densest gas element in the simulation is largely set by the equation of state, as can be seen from Figure~\ref{fig:adiabatic_track} where the different phases of the collapse are evident: First the isothermal collapse, then an  adiabatic rise of the temperature during the formation and gain in mass of the first core, followed by a relatively slow increase of temperature as the second collapse commences, and finally a rapid increase in temperature as the second core forms. 

We can see a decrease in the slope around 100 K as the rotational degrees of freedom of H$_2$ become excited and the effective adiabatic index drops from $\frac{5}{3}$ to $\frac{7}{5}$. Figure~\ref{fig:dens_over_time} shows the evolution of the maximum density over time (from the beginning of the simulation) from the point where the density has reached $10^{-15}$~g~cm$^{-3}$ to the end of the simulation, as well as the temperature of this densest fluid element at the same time. Less than 1.5~Kyr pass from reaching the threshold $10^{-15}$~g~cm$^{-3}$ to the collapse into the second core. The second collapse proceeds relatively much faster, as can be seen by the sharp rise in density and temperature once a density of  $10^{-8}$~g~cm$^{-3}$ has been reached. 

There are two notable points where the density decreases as a function of time, with a corresponding drop in temperature. They are also present (although at slightly different densities) in the non-ideal RMHD simulation by \cite{wurster2018collapse} and are also seen in the earlier work of \cite{bate2014sphmhd}. As these local maxima are also seen in barotropic calculations such as \cite{bate1998collapse} and \cite{mayer2025isolated}, this clearly cannot be a pure radiation effect. Instead, these represent phases where the first hydrostatic core `bounces' due to the stall of collapse as the equation of state hardens or rotational and magnetic forces  resist collapse, causing a slight outward expansion. The specific values of density where this effect occurs depend on the particular physics model and initial conditions used. As noted in the introduction, other gas will be hotter than the central element as it continues to be heated up by material collapsing in the centre. This is seen in Figure \ref{fig:scatterphases_nonideal}, which displays the gas and radiation temperature of every cell in the cloud core. The two are almost everywhere in equilibrium. Many cells are much hotter at a given density than the track followed by the densest gas element, and we observe that material almost down to $10^{-16}\,{\rm g\,cm^{-3}}$ has been heated above the initial temperature. This increase in temperature as compared to the adiabatic track has also been pointed out in the work of e.g. \cite{whitehouse2006rhd,tomida2013rmhd} and is the main effect that cannot be captured in a barotropic temperature treatment. The long neck in the middle of the panels is material with a soft equation of state,  which has a comparatively low opacity due to the significant drop in opacities at the evaporation of the dust grains, which causes an almost constant temperature in the low-opacity region. In between the two snapshots, the regions at densities $<10^{-11}$ g cm$^{-3}$ have continued to be heated up by the collapse of the gas in the centre. 

Figure~\ref{fig:non_ideal_early} shows different quantities in the simulation when the first core has already increased in density and temperature, and the second collapse has almost started. In the upper panels, we display the surroundings of the first core. As in earlier work \citep[e.g. ][]{tomida2013rmhd,bate2014sphmhd,vaytet2018birth,wurster2021nonidealimpactsingle,mayer2025isolated}, we observe magnetorotational outflows from the first core as material is moving away from the centre with a high rotational speed. As such, the azimuthal magnetic field is also strong in these outflows. The lower panels focus on the first core itself. The central magnetic field strength has reached more than 10 G, which is however still lower than simulations with ideal MHD at this stage of evolution \citep[e.g.][]{vaytet2018birth,wurster2021nonidealimpactsingle,mayer2025isolated}. Thermal ionization becomes efficient at much lower temperatures than the maximum temperature at this stage (from $\sim 600$~K on non-ideal MHD coefficients as obtained from {\small NICIL} drop rapidly) and as such the central region has `recoupled' to the magnetic field. 

Overall, the impact of non-ideal MHD with this specific choice of grain size distribution and cosmic-ray ionization rate appears to be somewhat reduced as compared to the barotropic run in \cite{mayer2025isolated}, which is unsurprising because with the simplification of the barotropic model, the recoupling happens at higher densities and, as noted above, temperatures are underestimated by the barotropic EOS. We observe material moving from the centre in the inner $\sim 3\,{\rm au}$ (visible in the radial velocity and as a jump in density and temperature in the form of a ring). This is connected to the drop in opacities at the dust evaporation, as radiation that was previously trapped in the centre can now suddenly escape more easily and heat up material sufficiently to exceed the dust evaporation temperature. As the dust evaporation front moves outwards, the pressure difference becomes strong enough to cause some gas to move outward. This is not seen in the simulations of \cite{tomida2013rmhd} and \cite{vaytet2013multigroup,vaytet2018birth}, which however employ different initial conditions and other, though similar, opacities. The fundamental process seems similar to the one causing the outflows observed by \cite{bate2011prestellardisks}, where material in the disk quickly becomes much hotter as it suddenly receives more radiation from the centre. We do not observe this behaviour for tests with constant opacity (but using the full equation of state), and we suspect this effect may become negligible when different opacities are employed (even just whether some smoothing is used in the table around this sharp drop). 

In Figure~\ref{fig:non_ideal_late}, we show a later evolutionary phase, namely when the second core has formed. The most prominent feature in the upper panels is the outflow from the second core, which is much faster than that from the first core with a radial velocity of $v_{\rm rad}>10$~km~s$^{-1}$ and features a more collimated structure. As such, this likely represents the onset of a protostellar jet. We also observe a rapidly rotating elongated structure, a `second disk' at scales $<0.5\,{\rm au}$, which is a phenomenon not observed in simulations with ideal MHD \citep[e.g. ][]{tomida2013rmhd,vaytet2018birth,ahmad2025birth}. However, while non-ideal MHD effects acting during earlier phases of the collapse have allowed sufficient angular momentum to reach the interior $1\,{\rm au}$ to form such a disk, this entire region is essentially in the ideal MHD limit due to high temperatures. This allows a strong amplification of the magnetic field as compared to the earlier snapshot (Figure \ref{fig:non_ideal_early}), up to more than $10^5\,{\rm G}$, from a combination of the dragging-in of the magnetic field during the second collapse and the winding-up due to rotation (as shown by the  azimuthal component). 

The lower panels of Fig.~\ref{fig:non_ideal_late} show the second core, clearly visible as a structure with very high density ($> 10^{-2}$~g~cm$^{-3}$) and temperature. The second core shows relatively little rotation and hosts the very strong magnetic field mentioned above. This field is consistent with the strength seen in the ideal MHD run of \cite{ahmad2025birth}, although higher than in the non-ideal MHD simulation. This result is somewhat expected, as their non-ideal MHD model \citep{marchand2016chemistry} has significantly stronger diffusion than ours, mainly due to the high cosmic-ray ionization rate we assume here. On the other hand, the field is stronger than that seen in all simulations run by \cite{wurster2018collapse}, even though some of these used much smaller cosmic-ray ionization rates (up to $10^{-12}\,{\rm s}^{-1}$). These differences are likely unrelated to the radiation treatment and instead caused by differences in initial conditions and MHD schemes. Indeed, based on the simulations in \cite{wurster2022originII}, it is likely that the maximum magnetic field strength is not converged at the resolution of \cite{wurster2018collapse}, and higher resolution results in stronger maximum magnetic fields over a wide range in mass resolutions.

Finally, we address the computational cost of our FLD implementation. This cost is heavily dependent on the number of Newton iterations taken for a single timestep. If a single iteration is done on all timebins, the cost of the implicit solve (which is essentially all of the cost of the FLD scheme) is about 25-30\% of the total cost in this specific simulation. If significantly more iterations are required (typically this is only the case for the highest timebin), the relative cost can increase substantially for a period of the simulation, up to 90\%, which also affects the total runtime. This motivates us to aim for improving our stopping criterion in future work. Even so, the non-ideal RMHD run presented here took less than 10000~cpu hours with 112 mpi-tasks. In addition, our next step is to perform a set of simulations with stronger non-ideal MHD effects, where smaller timesteps are taken in any case (aiding convergence) and the simulations will be more heavily dominated by the lower timebins (those with smaller timesteps), which generally converge more easily.

\section{Conclusions}
\label{sec:conclusions}
In this work we have presented an implementation of radiative transfer in the form of flux-limited diffusion for the moving-mesh code {\small AREPO}, accompanied by a physical model for the collapse of molecular cloud cores to protostars. We have benchmarked the accuracy of our code on standard test problems in the literature and found good agreement, which demonstrates the soundness of our numerical methods. This includes the version of our scheme utilizing local timestepping, which gives results that properly match those of global timesteps, despite the non-conservation of energy inherent to the use of Dirichlet boundaries. 

The scheme was applied to the problem of the collapse of cloud cores to protostars, and we have run the first non-ideal RMHD simulations with {\small AREPO}, producing results similar to those of previous studies. This demonstrates the capability of our implementation to handle complex and realistic astrophysical applications with a large range of densities and temperatures, and that it can be integrated with other modules of the code, such as non-ideal MHD.

In our collapse simulation with non-ideal RMHD, we observe both the launching of a fast-rotating (magneto-rotational) first core outflow as well as a fast (v$_{\rm rad}> 10\, {\rm km\,s^{-1}}$) and collimated jet from the second core. The second core is surrounded by a small disk (with radius $< 0.5$~au) enabled by the inclusion of non-ideal MHD. The magnetic field in the second core is amplified up to more than $10^5$~G, with a substantial azimuthal component. We plan to follow the evolution of the jet and stellar core for a longer amount of time in the future.

A natural extension of our method is to couple our machinery with the available cooling prescriptions, assuming chemical equilibrium \citep[akin to][which however also includes some non-equilibrium chemistry{}]{bate_keto2015ism}, adding source- and sink-terms to the equations for internal energy and radiation. For this, the use of an iteration over multiple implicit solves is a strict requirement, as the cooling times in diffuse regions are expected to be considerably smaller than the hydrodynamical timestep. 

\section*{Acknowledgements}
We thank the anonymous referee of this article for helpful comments which have improved this work. ACM thanks Matthew Bate for taking the time to explain the details of the FLD implementation in {\small SPHNG}, which was very helpful in extending the method to local timesteps. 
TN and PC acknowledge the support of the Deutsche Forschungsgemeinschaft (DFG, German Research Foundation) under Germany’s Excellence Strategy - EXC-2094 - 390783311 of the DFG Cluster of Excellence ``ORIGINS”. Support for OZ was provided by Harvard University through the Institute for Theory and Computation Fellowship.

\section*{Data Availability}
The data underlying this paper will be shared upon reasonable request to the corresponding author.

\bibliographystyle{mnras}
\bibliography{main.bib}

\label{lastpage}
\end{document}